\documentclass[usenatbib,useAMS]{mn2e}

\usepackage{color}
\usepackage{amsfonts}
\usepackage{amssymb}
\usepackage{mathtools}
\usepackage{relsize}
\usepackage{booktabs}
\usepackage{threeparttable}

\newcommand{\eg}{e.g.\ }
\newcommand{\Eg}{E.g.\ }
\newcommand{\ie}{i.e.\ }

\newcommand{\alert}[1]{\textcolor{black}{#1}}             % highlight text for later editing

\newenvironment{alertenv}{\color{black}}{\color{black}}
\newcommand{\vphi}{\ensuremath{v_\varphi}}              % azimuthal velocity
\newcommand{\vs}{\ensuremath{v_r}}                      % radial velocity
\newcommand{\cs}{\ensuremath{c_\mathsf{s,c}}}             % speed of sound
\newcommand{\sam}{\ensuremath{\ell}}                    % specific angular momentum
\newcommand{\avel}{\ensuremath{\Omega}}                 % angular velocity
\newcommand{\den}{\ensuremath{\varrho}}                 % density
\newcommand{\pres}{\ensuremath{P}}                      % pressure
\newcommand{\sden}{\ensuremath{\Sigma}}                 % surface density
\newcommand{\spre}{\ensuremath{\Pi}}                    % surface pressure
\newcommand{\Trphi}{\ensuremath{T_{r\varphi}}}          % stress tensor component
\newcommand{\kvis}{\ensuremath{\nu}}                    % kinematic viscosity
\newcommand{\cent}[1]{\ensuremath{#1_\mathsf{c}}}       % midplane quantities
\newcommand{\kepler}[1]{\ensuremath{#1_\mathsf{K}}}     % Keplerian quantities
\newcommand{\Mc}{\ensuremath{M_\star}}
\newcommand{\Msun}{\ensuremath{\mathrm{M}_\odot}}       % solar mass
\newcommand{\yr}{\ensuremath{\mathrm{yr}}}              % year
\newcommand{\di}[1]{\ensuremath{\mathsf{d}#1}}                        % integral measure
\newcommand{\dd}[2]{\ensuremath{\frac{\mathsf{d} #1}{\mathsf{d}#2}}}   % derivative
\newcommand{\del}[2]{\ensuremath{\frac{\partial #1}{\partial #2}}}     % partial derivative
\newcommand{\order}[1]{\ensuremath{\mathcal{O}\left(#1\right)}}        % order of magniture
\newcommand{\inv}[1]{\ensuremath{\frac{1}{#1}}}         % 1/x
\newcommand{\tinv}[1]{\ensuremath{\tfrac{1}{#1}}}       % tiny 1/
\title[Self-Gravitating Accretion Discs]%
	{Self-similar Evolution of Self-Gravitating Viscous Accretion Discs}

\author[T.\ F.\ Illenseer and W.\ J.\ Duschl]%
	{Tobias F.\ Illenseer$^{1}$\thanks{tillense@astrophysik.uni-kiel.de}
	and Wolfgang J.\ Duschl$^{1,2}$\thanks{wjd@astrophysik.uni-kiel.de}\\
	$^1$Institut f\"ur Theoretische Physik und Astrophysik, %
  Christian-Albrechts-Universit\"at zu Kiel, Leibnizstr. 15, 24118 Kiel, Germany \\
	$^2$Steward Observatory, The University of Arizona, 933 N. Cherry Ave., Tucson, AZ 85721, USA
	}

\begin{document}
\maketitle

%% abstract
\begin{abstract}
A new one-dimensional, dynamical model is proposed for geometrically thin,
self-gravitating viscous accretion discs. The vertically integrated equations
are simplified using the slow accretion limit and the monopole approximation
with a time-dependent central point mass to account for self-gravity and
accretion. It is shown that the system of partial differential equations can be
reduced to a single non-linear advection diffusion equation which describes the
time evolution  of angular velocity.

In order to solve the equation three different turbulent viscosity prescriptions
are considered. It is shown that for these parametrizations the differential equation
allows for similarity transformations depending only on a single non-dimensional
parameter. A detailed analysis of the similarity solutions reveals that this
parameter is the initial power law exponent of the angular velocity distribution
at large radii. The radial dependence of the self-similar solutions is in most
cases given by broken power laws. At small radii the rotation law always becomes
Keplerian with respect to the current central point mass. In the outer regions
the power law exponent of the rotation law deviates from the Keplerian value and
approaches asymptotically the value determined by the initial condition. It is
shown that accretion discs with flatter rotation laws at large radii yield
higher accretion rates.

The methods are applied to self-gravitating accretion discs in active galactic
nuclei. Fully self-gravitating discs are found to evolve faster than nearly
Keplerian discs. The implications on supermassive black hole formation and Quasar
evolution are discussed.

\end{abstract}

%% keywords

%% paper text starts here:
\section{Introduction}
\label{sec:introduction}

Accretion discs had become a major topic of astrophysical research during the
past five decades since the discovery of the first quasars in the early 1960's
\citep{matthews1963,schmidt1963} and their widely accepted theoretical explanation
by \cite{zeldovich1964}, \cite{salpeter1964} and \cite{lyndenbell1969}. Since
those days a variety of objects have been
identified in which mass accretion is the fundamental process. Among these are
such different objects as T-Tauri stars, cataclysmic variables (CV) and active
galactic nuclei (AGN) \citep{robinson1976,rees1984,appenzeller1989}. Thus a deep
understanding of the accretion process forms the basis for the explanation of
several important astrophysical processes from the formation of planetary systems
to the evolution of super-massive black holes (SMBHs) in galactic centres.

In the classical theory of disc accretion \citep{weizsaecker1948,luest1952,%
shakura1973,lyndenbell1974} viscous torques in the differentially rotating gas
flow around a central object cause redistribution of angular momentum from the
inner disc to outer regions. Thus the inner parts which are no longer fully
supported by centrifugal forces move inwards in the gravitational potential.
The same viscous stress transforms gravitational energy into heat leading to an
intense radiation.

The probably most crucial quantity regarding the accretion process is the
prescription of viscosity. Although we still lack a profound theory of the
underlying processes, there exists broad agreement that the nature of the viscosity
must be turbulent, \alert{at least in case of non self-gravitating discs}. 
This is strongly supported by a simple time-scale argument
already raised by \cite{goldreich1967} in connection with angular momentum
transport in rotating stars, which rules out any important contribution of
molecular viscosity \cite*[see][for an application to accretion discs]{frank2002}.
The perhaps most popular model of turbulent viscosity in this context was proposed
by \cite{shakura1973}. On the basis of their viscosity model they were able to
derive stationary solutions for geometrically thin accretion discs. In order to
solve the problem they implied that the gravitational potential of the central
object dominates thereby assuming a massless disc, which is in many cases a
fairly good approximation. However, at least in AGN accretion discs 
there is observational evidence that there exist discs with a non-Keplerian
rotation law \alert{\citep{greenhill1996,lodato2003,kondratko2005,hure2011}}.

Unfortunately, things become a lot more complicated if the mass distribution
within the disc contributes considerably to the overall gravitational potential.
In addition to the usual disc equations one has to solve Poisson's equation
for the gravitational potential introducing a non-linear coupling between mass
distribution and rotational velocity. The problem of a viscous self-gravitating
gaseous disc has already been tackled by \cite{weizsaecker1948} and
\cite{trefftz1952} who derived the basic differential equations. They discussed
some special solutions, but did not succeed in solving the general problem.

\alert{Another challenge when dealing with self-gravitating discs is their
ability to generate strong instabilities as was already pointed out by
\cite{toomre1964}. At first glance this seems quite desirable, because this
would lead to the proposed turbulence. \cite{laughlin1994} could indeed show
that simple one-dimensional diffusion models using $\alpha$-viscosity approximate
the main properties of three-dimensional simulations quite well. But their
analysis also reveals that the diffusive transport is less characterized by
a turbulent cascade and more through the action of gravitational torques. This
leads to a serious problem raised by \cite{balbus1999}. They argue that a
viscous parametrization may be inadequate because it can only describe the
dissipation of energy locally whereas gravitational forces can act over large
distances and are therefore non-local. In reply to this \cite{gammie2001} shows
in a seminal paper that in geometrically thin discs without large scale structures
the local treatment is applicable for simple cooling models where the cooling
time $\tau_\mathrm{c}$ is constant. He derives a simple formula relating
the $\alpha$ parameter of the \citeauthor{shakura1973} viscosity prescription to
$\Omega\tau_\mathrm{c}$. \citeauthor{gammie2001} furthermore shows that
self-gravitating accretion discs fragment if the cooling time falls below a
critical value of $3\Omega^{-1}$ \cite[see also][]{mejia2005}.}

\alert{His findings are based on the analysis of two-dimensional
shearing box simulations and were later confirmed by \cite{rice2003} and
\cite{lodato2004} in global three-dimensional SPH simulations. In case of
more massive discs the situation seems more complex. \cite{lodato2005}
report no clear evidence for global transport of energy induced by gravitational
forces if the aspect ratio $H/r<1/10$. This was later confirmed by \cite{cossins2009},
but they show that for mass ratios of $M_\mathrm{disc}/M_\star=0.125$ the fraction
of non-local wave energy transport rises up to $15\,\%$. If $M_\mathrm{disc}/M_\star$
exceeds $1/2$ global transport dominates \citep{forgan2011}.
Contrary to these results global grid based simulations
show that even in the low mass regime local $\alpha$-models may not be applicable
at all \citep{mejia2005} especially when considering more realistic cooling models
\citep{boley2006}. Recent grid based simulations \citep{michael2012,steiman2013}
however show that averaging the results over many dynamic times and large spatial
volumes yields roughly the radial dependence of the $\alpha$ parameter predicted
by \cite{gammie2001}. Thus a local viscous approximation seems applicable as
long as the disc is not too thick and too massive.}

\alert{We would like to emphasize that
most of the results reviewed in the preceding paragraphs were obtained for
protoplanetary discs where the aspect ratio becomes $1/10$ even for moderate
disc masses. In case of AGN discs the situation is somewhat different, because
the aspect ratios in these discs are thought to be smaller than those found in
protoplanetary discs by roughly an order of magnitude \citep{collin1990,lin1996}.
Since the aspect ratio has a major impact on global wave transport \citep{lodato2004}
one may model AGN discs with local models for mass ratios well above one.}

The modern treatment of self-gravitating accretion discs \alert{using simple
one-dimensional models} begins with the work of \cite{paczynski1978} who
solves -- under certain assumptions -- the vertical structure
problem for geometrically thin discs. He also introduces the concept of self-regulation
which is based on the idea that radiative cooling and viscous heating adjust the
temperature in a way that keeps the disc in a marginally stable state. Thereby
he assumes that turbulence is driven by gravitational instabilities. 
\cite{sakimoto1981} modify this work by replacing the turbulence model with the
$\alpha$-parametrization of \cite{shakura1973}. They derive stationary solutions
and apply them to self-gravitating AGN accretion discs.

Based on these early attempts to construct self-gravitating accretion disc
models \cite{mineshige1996} and \cite{bertin1997} found stationary self-similar
solutions. The latter author combines the $\alpha$-viscosity model with the
concept of self-regulation \citep[see also][]{bertin1999}. Thereby he assumes that
the disc is always in an marginally stable state, \ie the \citeauthor{toomre1964} parameter
\citep{toomre1964} is close to unity. With help of this assumption he avoids the
problem of solving the energy equation to determine the temperature of the disc 
and thus the speed of sound which is necessary to compute $\alpha$-viscosity
(see Sec.~\ref{sec:viscosity_prescription}).

\alert{\cite{mineshige1997}} and \cite{tsuribe1999} develop self-similar time-dependent
solutions \alert{based on} the $\alpha$-prescription. The former authors additionally
assume that the pressure scale height of the disc scales linearly with radius.
\alert{This assumption modifies the viscosity prescription considerably because
it scales with sound speed multiplied by radius instead of scale height in contrast
to the original $\alpha$-viscosity. In both papers the discs are isothermal and self-gravity is
treated in the monopole approximation. \cite{mineshige1997a} propose a non-isothermal
model by introducing a polytropic relation to account for a fixed radial temperature gradient.}
They also discuss the possibility of Quasar formation on the basis of their model. All these
results seem quite promising, but the time scales for AGN evolution always exceed
the Hubble time for reasonable disc parameters as was already pointed out by \cite*{shlosman1990}.
Another problem of stationary self-gravitating $\alpha$-discs was pointed out by
\cite*{duschl2000} who demonstrate that one inevitably yields temperature distributions
which are inconsistent with the thin disc assumption.

\cite{lin1987} propose a new viscosity prescription where the effective kinematic
viscosity is determined by the typical length scale of unstable regions. They
show that in a gravitationally unstable disc this length scale depends on the
local mass distribution and the rotational velocity. With help of their new
viscosity model, they derive time-dependent self-similar solutions. However these
discs cannot be considered as fully self-gravitating, because the authors keep
the central mass constant in their models and assume Keplerian rotation.
Another problem of these solutions is that the viscosity model requires
gravitational instabilities to generate a sufficiently high effective viscosity.
Hence their effective viscosity tends to zero in the Keplerian limit, because
Keplerian discs are known to be stable \citep{safronov1958}.

In a completely different approach the effective viscosity is coupled to the
critical Reynolds number \citep{duschl1998,richard1999}. This so called 
$\beta$-prescription relies on the observation that in laboratory experiments
almost any flow becomes turbulent at high Reynolds numbers. Since the Reynolds
numbers in accretions discs are extraordinary high, one can usually expect that
these flows are turbulent regardless of the actual mechanism that generates the
turbulence \citep{luest1952}. On the basis of this new viscosity prescription
\cite{duschl2000} develop stationary solutions for self-gravitating accretion
discs. In this context they also discuss the possibility of supermassive black
hole formation based on their model.

\cite*{abbassi2006} and more recently \cite*{abbassi2013} derive self-similar
solutions for self-gravitating $\beta$-viscous discs. Their models modify those
for polytropic discs of \cite*{mineshige1997a} by replacing the $\alpha$-prescription
and therefore avoid the drawbacks discussed above. However, by doing so they
encounter a problem not further dealt with by the authors. In order to derive
the self-similar solution they introduce a similarity variable which depends on
the proportionality constant $K$ \alert{and the exponent $\gamma$} of the polytropic relation $\pres=K\den^\gamma$.
\alert{Both constants enter} the set of differential equations only due to
the pressure gradient in the radial momentum equation. As we will show in
Sec.~\ref{sec:slow_accretion_limit}, this term is usually negligible and the
authors do actually neglect it by using the slow accretion limit. Thus, although
the \alert{parameters $K$ and $\gamma$ are removed from the underlying basic equations, their
similarity solutions depend on them which causes a serious contradiction between
model assumptions and solutions.}

Therefore, in the present paper we simplify the set of differential equations
before applying an appropriate similarity transformation. This approach yields a
single partial differential equation (PDE) for self-gravitating accretion disc
dynamics (Sec.~\ref{sec:disc_equation}). Although our general derivation is
independent of the viscosity prescription one has to select a specific model in
order to obtain solutions of the differential equation. Therefore we discuss
three different viscosity models including the $\beta$-viscosity
(Sec.~\ref{sec:viscosity_prescription}). Furthermore we show that for these
viscosity models our disc evolution equation is invariant under the same scaling
transformation which admits a similarity transformation depending on a single
non-dimensional parameter $\kappa$ (Sec.~\ref{sec:differential_equation}). Thus
we obtain a hole family of self-similar solutions, each with a different value
of $\kappa$. We demonstrate that this parameter is related to the slope of the
rotational velocity far from the origin. In addition we show that $\kappa$ has
fundamental impact on the evolution of self-gravitating discs and that discs
with an asymptotically flatter rotation law have higher accretion rates and
therefore evolve faster than those with a nearly Keplerian rotation law
(Sec.~\ref{sec:impact_of_kappa}).

\section{The disc model}
\label{sec:disc_model}

Our model is based on the standard theory of geometrically thin, axisymmetric
accretion discs according to \cite{weizsaecker1948} and \cite{luest1952}
\cite[for a modern treatment, see][]{kato2008}. According to this we assume
that the disc is in hydrostatic balance in the vertical direction. This allows
us to decouple the dynamical evolution from the vertical structure equations by
introducing the vertically integrated density
\begin{equation}
  \label{eqn:surface_density}
  \sden(t,r) = \int_{-H}^H \den(t,r,z)\di{z}
\end{equation}
and pressure
\begin{equation}
  \label{eqn:surface_pressure}
  \spre(t,r) = \int_{-H}^H \pres(t,r,z)\di{z}.
\end{equation}
Thereby the limits of integration are given by the so far unspecified parameter
$H$ which can be finite or infinite\footnote{For $H\to\infty$, our definition of
$\sden$ agrees with the usually used version}. We show in Sec.~\ref{sec:vertical_structure}
that $H$ is related to the pressure scale height. It is important to mention here,
that even in the case where $H$ becomes large the vertical density and pressure
gradients are rather steep, so that the thin disc assumption always holds. In
addition to surface density and integrated pressure we define the vertically
integrated $r$-$\varphi$-component of the stress tensor:
\begin{equation}
  \label{eqn:stress_tensor}
  \Trphi(t,r) = \int_{-H}^H t_{r\varphi}(t,r,z)\di{z}
  = \kvis\sden r\del{\avel}{r}
\end{equation}
which is usually the dominant term. $\avel=\vphi/r$ is the angular velocity
and $\kvis$ the kinematic viscosity,
both are assumed to be independent of the vertical coordinate $z$ in order to
carry out the integration. The set of differential equations we consider in this
work are then given by the continuity equation
\begin{equation}
  \label{eqn:continuity}
  \del{\sden}{t} + \inv{r}\del{}{r}\Bigl(r\vs\sden\Bigr) = 0
\end{equation}
and the transport equations for radial momentum
\begin{equation}
  \label{eqn:radial_momentum}
  \del{\vs}{t} + \vs\del{\vs}{r} = -\inv{\sden}\del{\spre}{r} + \frac{\vphi^2}{r} - \del{\Phi}{r}
\end{equation}
and angular momentum
\begin{equation}
  \label{eqn:angular_momentum}
  \del{\sam}{t} + \vs\del{\sam}{r} = \inv{r\sden}\del{}{r}\Bigl(r^2\Trphi\Bigr)
\end{equation}
where $\Phi$ is the gravitational potential, $\sam=r\vphi=r^2\Omega$ is the
specific angular momentum and $\Trphi$ is given by Eq.~(\ref{eqn:stress_tensor}).
In order to solve the system above one has to consider the vertical balance law.
We derive an approximate solution of the vertical structure equation for an
ideal gas equation of state assuming a polytropic relation between pressure
and density in Sec.~\ref{sec:vertical_structure}. In addition one generally has
to solve some kind of energy transport equation to determine the thermodynamic
structure of the disc. However, we show in Sec.~\ref{sec:slow_accretion_limit}
that for geometrically thin discs the system decouples from the energy equation
if all radial gradients are moderate and the viscosity $\nu$
does not depend on temperature. In Sec.~\ref{sec:viscosity_prescription} we
discuss some reasonable viscosity prescriptions with this property.

Since the discs we examine in this work are assumed to be self-gravitating, we
have to solve Poisson's equation to obtain the gravitational acceleration
$\del{\Phi}{r}$ in Eq.~(\ref{eqn:radial_momentum}). This is done using the
monopole approximation (see Sec.~\ref{sec:monopole_approximation}).

\subsection{Vertical structure}
\label{sec:vertical_structure}

The derivation of the vertical structure equations generalizes the work of
\cite{hoshi1977} who studied polytropic Keplerian discs and \cite{paczynski1978}
who also included the discs potential, but in a slightly different way than we do.

The basic assumptions are hydrostatic equilibrium between pressure forces and
gravitational forces
\begin{equation}
  \label{eqn:hydrostaic_equilibrium}
  \inv{\den}\del{\pres}{z} = -\del{\Phi}{z}
\end{equation}
and a polytropic relation according to
\begin{equation}
  \label{eqn:polytropic_relation}
  \pres = K \den^\frac{n+1}{n},\quad 0<K,\quad 0<n.
\end{equation}
The constant $K$ can be determined from the midplane values of density and
pressure
\begin{equation*}
  K = \cent{\pres}\cent{\den}^{-\frac{n+1}{n}} = \tfrac{n}{n+1}\cs^2\cent{\den}^{-\inv{n}}
\end{equation*}
where $\cs$ is the midplane polytropic sound velocity
\begin{equation}
  \label{eqn:midplane_sound_velocity}
  \cs^2 = \biggl(\dd{\pres}{\den}\biggr)_{z=0} = \tfrac{n+1}{n} K \cent{\den}^{\inv{n}}
  = \tfrac{n+1}{n} \frac{\cent{\pres}}{\cent{\den}}.
\end{equation}
The polytropic relation allows us to express the left hand side of
Eq.~(\ref{eqn:hydrostaic_equilibrium}) in terms of $\den$ alone:
\begin{equation*}
  \inv{\den}\del{\pres}{z}=n\cs^2\del{}{z}\biggl(\frac{\den}{\cent{\den}}\biggr)^\inv{n}
  = -\del{\Phi}{z}.
\end{equation*}
This differential equation can be integrated immediately using the boundary conditions
$\den(r,z=0)=\cent{\den}(r)$ and $\Phi(r,z=0)=\cent{\Phi}(r)$:
\begin{equation}
  \label{eqn:vertical_density_stratification_exact}
  \den(r,z) = \cent{\den}\,\biggl(1-\frac{\Phi-\cent{\Phi}}{n\cs^2}\biggr)^n
\end{equation}
where the midplane values of density $\cent{\den}$, speed of sound $\cs$ and
gravitational potential $\cent{\Phi}$ are functions of radius $r$. The polytropic
index $n$ may also depend on $r$. In contrast to the solution given in
\cite{paczynski1978} who replaced the gravitational acceleration in the
vertical balance law (\ref{eqn:hydrostaic_equilibrium}) by an approximate 
solution of Poisson's equation for the gravitational potential, our result in
Eq.~(\ref{eqn:vertical_density_stratification_exact}) is exact. However, since
the gravitational potential $\Phi$ depends on $\den$ in self-gravitating discs,
we cannot deduce the density profile in terms of analytic functions in general.

Nevertheless, if we are dealing with thin discs, we can approximate the value
of the gravitational potential using Taylor expansion around the midplane
\citep{luest1952}
\begin{equation*}
  \Phi(r,z) = \cent{\Phi}+z\biggl.\del{\Phi}{z}\biggr|_{z=0}
    +\frac{z^2}{2}\biggl.\del{^2\Phi}{z^2}\biggr|_{z=0}+\order{z^3}.
\end{equation*}
If we furthermore assume that the potential is symmetric with respect to the
midplane the linear and cubic terms vanish and this approximation is of order
$z^4$. Inserting this expansion into Eq.~(\ref{eqn:vertical_density_stratification_exact})
yields an approximate expression of vertical density stratification for
potentials with mirror symmetry:
\begin{equation}
  \label{eqn:vertical_density_stratification}
  \den(r,z)=\cent{\den}\Biggl(1-\inv{2n}\biggl(\frac{z}{h}\biggr)^2\Biggr)^n
\end{equation}
with the scale height
\begin{equation}
  \label{eqn:scale_height}
  h(r)=\cs\sqrt{\del{^2\Phi}{z^2}\biggr|}_{z=0}^{~-1}.
\end{equation}
In addition to that we define the geometric height or half-thickness of the disc
as the vertical extend at which the density vanishes:
\begin{equation}
  \label{eqn:disc_height}
  H(r) = \sqrt{2\,n}~h(r).
\end{equation}
This definition of $H$ fixes the integration limits in Eqs.~(\ref{eqn:surface_density})
and (\ref{eqn:surface_pressure}). In the isothermal limit $n\to\infty$ we
therefore yield $H\to\infty$ and Eq.~(\ref{eqn:vertical_density_stratification})
becomes the well known Gaussian profile \citep{lyndenbell1969,shakura1973}
\begin{equation}
  \label{eqn:vertical_density_stratification_isotherm}
  \den_\mathsf{isoth}(r,z)=\cent{\den}e^{-\inv{2}\left(\frac{z}{h}\right)^2}.
\end{equation}
Furthermore we may utilize the definition of $H$ and the vertical density
stratification (\ref{eqn:vertical_density_stratification}) to compute the surface
density and with help of the polytropic relation (\ref{eqn:polytropic_relation})
the integrated pressure
\begin{align}
  \label{eqn:surface_density_integrated}
  \sden &= \cent{\den}H \sqrt{\pi}
     \frac{\Gamma(n+1)}{\Gamma\left(n+\frac{3}{2}\right)} \\
  \label{eqn:surface_pressure_integrated}
  \spre &=  \cent{\pres}H \sqrt{\pi}
     \frac{(n+1)~\Gamma(n+1)}{\left(n+\frac{3}{2}\right)\Gamma\left(n+\frac{3}{2}\right)}
\end{align}
where $\Gamma$ represents the gamma function. These relations have already been
derived by \cite{hoshi1977} in case of non self-gravitating discs. Dividing
Eq.~(\ref{eqn:surface_pressure_integrated}) by Eq.~(\ref{eqn:surface_density_integrated})
and using the expression for the midplane sound velocity (\ref{eqn:midplane_sound_velocity})
one obtains the useful relation
\begin{equation}
  \label{eqn:vertical_structure_solution}
  \spre = \eta \cs^2 \sden
\end{equation}
between integrated pressure, midplane speed of sound and surface density. The
non-dimensional function
\begin{equation}
  \label{eqn:definition_eta}
  \eta=\frac{n}{n+\frac{3}{2}}.
\end{equation}
is always larger than $0$ and becomes at most $1$ in the isothermal limit. Since
$\eta$ depends on the local polytropic index $n$ it will in general depend on
the radial coordinate $r$.

\begin{figure}
  \centering
  \includegraphics[width=\linewidth]{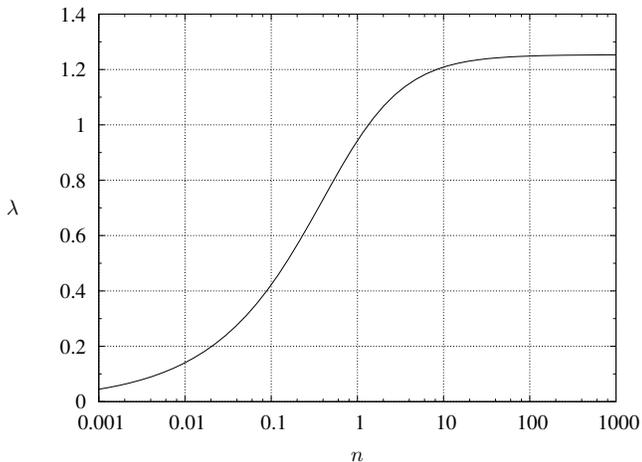}
  \caption{The non-dimensional factor $\lambda$ as a function of polytropic
    index $n$.}
  \label{fig:lambda_solution}
\end{figure}

With help of Eq.~(\ref{eqn:surface_density_integrated}) and
Eq.~(\ref{eqn:disc_height}) we can derive another useful equation which relates
surface density to central density and scale height:
\begin{equation}
  \label{eqn:surface_density_integrated2}
  \sden = 2\lambda\cent{\den}h.
\end{equation}
The non-dimensional factor
\begin{equation}
  \label{eqn:definition_lambda}
  \lambda = \sqrt{\frac{\pi n}{2}}
     \frac{\Gamma(n+1)}{\Gamma\left(n+\frac{3}{2}\right)}
\end{equation}
is shown in Fig.~\ref{fig:lambda_solution} as a function of polytropic index $n$.
For reasonable values of $n\gtrsim\tfrac{1}{2}$ it is roughly of order one and
becomes at most $\sqrt{\pi/2}$ in the isothermal limit \citep{luest1952,lyndenbell1969}.

A nice feature of the vertical structure equations given above is the possibility
to compute the scale height of different contributions to the potential and 
combine them to one effective scale height:
\begin{equation*}
  h = \sqrt{\sum_i \inv{h_i^2}}^{~-1}.
\end{equation*}
We demonstrate how to apply this to some important examples and compare the
results to those given in the literature.
\begin{enumerate}
  \item point mass $\Mc$:
    \begin{equation*}
      \Phi_\ast=\frac{G\,\Mc}{\sqrt{r^2+z^2}},\qquad
      \del{^2\Phi_\ast}{z^2}\biggr|_{z=0}=\frac{G\,\Mc}{r^3}=\kepler{\Omega}^2
    \end{equation*}
    where $\kepler{\Omega}$ is the Keplerian angular velocity. Hence the scale
    height of Keplerian discs is given by
    \begin{equation}
      \label{eqn:scale_height_kepler}
      h = \frac{\cs}{\kepler{\Omega}}.
    \end{equation}
    This result has been known for decades in case of isothermal discs 
    \citep[see, \eg][]{pringle1981} and was also obtained for polytropic discs
    by \cite{hoshi1977} and \cite{paczynski1978}.
  \item self-gravitating homogeneous slab:\\
    From Poisson's equation we have \citep[see, \eg][]{mestel1963}
    \begin{equation*}
      \del{^2\Phi}{z^2}\biggl|_{z=0}=4\pi G\cent{\den}
    \end{equation*}
    and therefore the scale height of the slab becomes
    \begin{equation}
      \label{eqn:scale_height_slab}
      h=\cs\sqrt{4\pi G\cent{\den}}^{~-1}.
    \end{equation}
    The half-thickness computed from this result by use of Eq.~(\ref{eqn:disc_height})
    differs from that given in \cite{paczynski1978} for polytropic 
    self-gravitating sheets. The deviation is of order one as long as $n$ is of
    order one and becomes large in the isothermal limit $n\to\infty$. However,
    in that case our solution reproduces exactly the result derived by
    \cite{spitzer1942}. 
  \item point mass and self-gravitating homogeneous slab:
    Combining (i) and (ii) as described above leads to the result proposed by
    \cite{luest1952} \cite[see also][]{sakimoto1981} who derived the scale height
    for self-gravitating isothermal sheets with central point mass:
    \begin{equation}
      \label{eqn:scale_height_pmslab}
      h = \cs\sqrt{4\pi G\cent{\den}+\kepler{\Omega}^2}^{~-1}.
    \end{equation}
    Thus again our more general result is consistent with the isothermal limit.
  \item self-gravitating axisymmetric disc within an axisymmetric external
    potential $\Phi_\mathsf{ext}$ in radial balance:
    \begin{equation}
      \label{eqn:scale_height_balance}
      h = \cs\Bigl(4\pi G\cent{\den}
        -2\Omega^2(1+x)+\Delta_{rz}\Phi_\mathsf{ext}\bigl|_{z=0}\Bigr)^{-\inv{2}}
    \end{equation}
    where $x$ is the logarithmic derivative of midplane angular velocity $\Omega$
    with respect to $r$ 
    \begin{equation}
      \label{eqn:power_law_exponent}
      x = \del{\ln\Omega}{\ln r}.
    \end{equation}
    and $\Delta_{rz}$
    is the axisymmetric Laplacian. The derivation uses again Poisson's equation
    for the disc potential and in addition the gravitational balance law
    (Eq.~\ref{eqn:centrifugal_gravity_balance}). This result has been derived by
    \cite{bertin1999} for isothermal discs. The contribution due to the external
    potential vanishes for all $r>0$ if $\Phi_\mathsf{ext}$ is the point mass
    potential. In that case the non self-gravitating limit ($\cent{\den}\to 0,
    x\to -\tfrac{3}{2}$) approaches the Keplerian value of the scale height.
    Another interesting case is \emph{Mestel's disc} \cite[see][]{mestel1963}
    which has $x=-1$ and the scale height becomes that of an infinite slab
    (\ref{eqn:scale_height_slab}).
\end{enumerate}

\subsection{The slow accretion limit}
\label{sec:slow_accretion_limit}

In this section we will introduce the slow accretion limit \citep{luest1952}
and show that the transport equation for radial momentum
(Eq.~\ref{eqn:radial_momentum}) simplifies to a balance law equating centrifugal
and gravitational forces. Thereby we make use of some relations already derived
in Sec.~\ref{sec:vertical_structure}. The basis for the derivation shown below
is the assumption that the rotational velocity of geometrically thin discs is highly
supersonic and that the radial drift velocity is subsonic, \ie $\vs\leq\cs\ll\vphi$
\citep[see, \eg][]{pringle1981}. In case of non self-gravitating discs
the requirement of supersonic rotation is quite obvious and a direct consequence
of the thin disc assumption. With help of Eq.~(\ref{eqn:scale_height_kepler}) one
concludes
\begin{equation}
  \label{eqn:supersonic_rotation_nsg}
  1 \ll \frac{r}{h} = \frac{r\Omega}{\cs} = \frac{\vphi}{\cs}.
\end{equation}
We cannot derive a similar estimate if we take self-gravity into account,
because in that case the scale height depends on mass distribution which in
turn influences the rotational velocity. This makes the relation between $h$ and
$\vphi$ more difficult. However, we will show in Section \ref{sec:supersonic_rotation}
that our approximations are at least consistent with the assumption given above.

With help of the vertical structure equation (\ref{eqn:vertical_structure_solution})
we can eliminate $\spre$ from the radial momentum equation (\ref{eqn:radial_momentum}).
After multiplication with $r/\cs^2$ one obtains
\begin{equation}
	\label{eqn:radial_momenum_transformed}
	\begin{split}
    \frac{r}{\cs^2}\del{\vs}{t} + \biggl(\frac{\vs}{\cs}\biggr)^2\del{\ln\vs}{\ln r}
%     = \biggl(\frac{\vphi}{\cs}\biggr)^2 - \frac{r}{\cs^2}\del{\Phi}{r} \\
%     -\eta\biggl\{\del{\ln\eta}{\ln r}+\del{\ln\cs^2}{\ln r}+\del{\ln\sden}{\ln r}\biggr\}
    = \biggl(\frac{\vphi}{\cs}\biggr)^2 - \frac{r}{\cs^2}\del{\Phi}{r} \\
    -\eta\biggl\{\frac{\tfrac{3}{2}}{n+\tfrac{3}{2}}\del{\ln n}{\ln r}
    +\del{\ln\cs^2}{\ln r}+\del{\ln\sden}{\ln r}\biggr\}.
  \end{split}
\end{equation}
Thereby we used Eq.~(\ref{eqn:definition_eta}) to express the logarithmic
derivative of $\eta$ in terms of the logarithmic derivative of the polytropic
index $n$. We can henceforth conclude that if the radial gradients of $n$, $\cs$
and $\sden$ are moderate, \ie their logarithmic derivatives are at most of order
one, we can neglect the terms within the curly brackets multiplied
by $\eta\leq 1$ in Eq.~(\ref{eqn:radial_momenum_transformed}) in comparison with
the term $(\vphi/\cs)^2\gg 1$. The same argument applies to the second term on
the left hand side, which is also of order one as long as the radial drift velocity
is subsonic. The remaining terms of the radial momentum equation are
\begin{equation*}
  r\del{\vs}{t} = \vphi^2 - r\del{\Phi}{r}.
\end{equation*}
In the slow accretion limit one expects that temporal changes of the radial 
drift velocity $\vs$ occur on the viscous time-scale
$\tau_\mathsf{vis}\gg\tau_\mathsf{dyn}\approx\Omega^{-1}=r/\vphi$. Hence we may
approximate the term on the left hand side by
\begin{equation*}
  r\del{\vs}{t}\approx r\frac{\vs}{\tau_\mathsf{vis}}
  = \vs\vphi\frac{\tau_\mathsf{dyn}}{\tau_\mathsf{vis}} \ll \vphi^2
\end{equation*}
and the radial momentum transport equation reduces to the gravitational balance
law
\begin{equation}
  \label{eqn:centrifugal_gravity_balance}
  \vphi^2 = r\del{\Phi}{r}.
\end{equation}
This result has been known for decades since the early works of \cite{weizsaecker1948}
and \cite{luest1952}. However, to our knowledge it has never been derived in such
a general way using the vertical structure equations to rewrite the radial
pressure gradient. By doing so we can explicitly show that radial pressure forces
in geometrically thin discs are usually rather small compared to centrifugal
and gravitational forces.

\subsection{Monopole approximation and self-gravity}
\label{sec:monopole_approximation}

So far we did not point out how to compute the radial gravitational acceleration
$-\del{\Phi}{r}$. In general there would be contributions from the central object
and the mass distribution within the disc. Thus it would involve the solution of
Poisson's equation which is difficult even in case of rotationally symmetric and
geometrically thin systems. In this section we will introduce the monopole
approximation for such systems and derive an approximate solution to
Eq.~(\ref{eqn:centrifugal_gravity_balance}).

The derivation basically follows the method of \cite{toomre1963} who uses
Hankel transforms to compute the potential of razor-thin discs given in the
equatorial plane by
\begin{equation}
  \label{eqn:poisson_solution}
  \Phi_\mathrm{d}(r) = -2\pi G\int\limits_0^{\infty} \di{k} J_0(kr)
		\int\limits_0^{\infty}\sden(s)J_0(ks)s\di{s}.
\end{equation}
Here $J_0$ denotes the Bessel function of the first kind of order $0$.
At about the same time \cite{mestel1963} found that the radial gravitational
acceleration $-\del{\Phi}{r}$ at a certain distance from the centre $r$ can be
split up into three parts: The monopole term which depends on the enclosed 
mass (see Eq.~\ref{eqn:enclosed_mass}) and contributions from the mass within
and beyond $r$ \citep[see also][]{mineshige1997}. This solution
involves spatial integrals over the mass distribution $\sden(r)$ which cannot be
evaluated analytically in general.

However for certain centrally condensed mass distributions the monopole term is
dominant and the two other terms cancel out as was already pointed out by
\cite{mestel1963}.
Hence one may neglect all terms except for the monopole term and approximate
the solution to Eq.~(\ref{eqn:centrifugal_gravity_balance}) by
\begin{equation}
  \label{eqn:monopole_approximation}
  \vphi^2 = \frac{G M(r)}{r}
\end{equation}
with the enclosed mass
\begin{equation}
  \label{eqn:enclosed_mass}
  M(r) = \Mc + 2\pi\int_0^r\sden(s)s\di{s}
\end{equation}
where $\Mc$ is the mass of the central object. Since the error introduced by
neglecting the two additional contributions depends on $\sden(r)$ it is difficult
to estimate it in general.

In order to elucidate this we show how to derive Eq.~(\ref{eqn:monopole_approximation})
from Eq.~(\ref{eqn:poisson_solution}). Unlike \cite{mestel1963} and \cite{mineshige1997} 
we do not split up the gravitational acceleration obtained from (\ref{eqn:poisson_solution})
via differentiation. Instead we take the solution (\ref{eqn:poisson_solution}) add
the potential of the central point mass and insert it into
Eq.~(\ref{eqn:centrifugal_gravity_balance}). Then we apply the inverse Hankel
transform which provides us with an integral expression of the surface density
\citep[see][]{binney1987}
\begin{equation*}
  \sden(r)=\inv{2\pi G}\int\limits_0^{\infty}\di{k} J_0(kr)
     \int\limits_0^{\infty}\biggl(\vphi(s)^2-\frac{G\Mc}{s}\biggr)
%        \del{}{s}\Bigl(J_0(ks)\Bigr)\di{s}.
        k J_1(ks) \di{s}.
\end{equation*}
If we insert this result in Eq.~(\ref{eqn:enclosed_mass}), we can compute
the enclosed mass. Thereby the contribution due to the central point mass within
the integral cancels the constant term $\Mc$. Thus we proceed utilizing
integration by parts with respect to $s$ considering that
\begin{equation*}
  \del{}{s}\Bigl(J_0(ks)\Bigr) = -k J_1(ks)
\end{equation*}
which gives us
\begin{equation*}
  \begin{split}
  M(r) = \frac{r}{G}\biggl\{
     \int\limits_0^\infty\del{\vphi^2}{s}\mathcal{F}(r,s)\di{s}
     -\Bigl[\vphi(s)^2\mathcal{F}(r,s)\Bigr]_{s=0}^{s=\infty}\biggr\}
  \end{split}
\end{equation*}
with
\begin{equation}
	\label{eqn:Fltgt_functions}
  \mathcal{F}(r,s)=\int\limits_0^\infty\inv{k}J_1(kr)J_0(ks)\di{k}
  = \begin{cases}
      \mathcal{F}_<(\tfrac{s}{r})\ \mathrm{if}\ s\leq r\\
      \mathcal{F}_>(\tfrac{r}{s})\ \mathrm{if}\ s>r\\
    \end{cases}.
\end{equation}
The solution of this definite integral depends on whether $r<s$ or not. An
analytical expression in terms of complete elliptic integrals is given in
App.~\ref{app:weight_functions} together with asymptotic expansions which allow
us to evaluate the surface terms. We may now split the integral into an
inner $s<r$ and outer $s>r$ part
\begin{equation*}
  \begin{split}
	\frac{GM(r)}{r} &=\ \vphi(0)^2
		-\tinv{2}\lim_{s\to\infty}\tfrac{r}{s}\vphi(s)^2 \\
	&+\int\limits_0^r\del{\vphi^2}{s}\mathcal{F}_<\bigl(\tfrac{s}{r}\bigr)\di{s}
	+ \int\limits_r^\infty\del{\vphi^2}{s}\mathcal{F}_>\bigl(\tfrac{r}{s}\bigr)\di{s}
	\end{split}
\end{equation*}
and substitute the variable of integration
\begin{equation*}
  \begin{split}
	&\frac{GM(r)}{r} =\ \vphi(0)^2
		-\tinv{2}\lim_{s\to\infty}\tfrac{r}{s}\vphi(s)^2 \\
	&+\int\limits_0^1\del{}{k}\Bigl(\vphi(rk)\Bigr)^2\mathcal{F}_<(k)\di{k}
	- \int\limits_0^1\del{}{k}\Bigl(\vphi(\tfrac{r}{k})\Bigr)^2\mathcal{F}_>(k)\di{k}.
	\end{split}
\end{equation*}
If we add $0=+1-1$ to the function $\mathcal{F}_<(k)$ in the first integral,
carry out one integration and factor out the term $\vphi(r)^2$ we finally get
the result
\begin{equation}
  \label{eqn:poisson_solution2}
  \begin{split}
	&\frac{GM(r)}{r} =\ \vphi(r)^2\Biggl\{1
		-\tinv{2}\lim_{s\to\infty}\frac{r}{s}\biggl(\frac{\vphi(s)}{\vphi(r)}\biggr)^2 \\
	& -\int\limits_0^1\mathcal{H}(r,kr)\frac{1-\mathcal{F}_<(k)}{k}\di{k}
	  +\int\limits_0^1\mathcal{H}\bigl(r,\tfrac{r}{k}\bigr)\frac{\mathcal{F}_>(k)}{k}\di{k}
	\Biggr\}
	\end{split}
\end{equation}
with
\begin{equation}
	\label{eqn:Hvphi}
  \mathcal{H}(r,s)=\frac{\vphi(s)^2}{\vphi(r)^2}\,\dd{\ln\vphi^2}{\ln s\ }
    = 2\frac{s^2\Omega(s)^2}{r^2\Omega(r)^2}\biggl(\dd{\ln\Omega}{\ln s\ }+1\biggr)
\end{equation}
The comparison with Eq.~(\ref{eqn:monopole_approximation}) reveals that the
exact solution of the thin disc Poisson problem reduces to the monopole
approximation if the sum in the curly brackets of
Eq.~(\ref{eqn:poisson_solution2}) is close to one. This sum only depends on the
rotation law given by $\Omega$ as a function of radial distance to the origin.
Thus we may ask ourselves if there exist certain rotation laws for which this
condition is fulfilled.

First of all one should restrict the discussion to rotation laws for which the
centrifugal acceleration $r\Omega^2$ tends to zero at infinity. Thus
$\Omega\propto r^\kappa$ with $\kappa<-\tinv{2}$ must hold in the limit
$r\to\infty$. Then we can neglect the surface term and the deviation from the
monopole approximation is completely determined by the two definite integrals
which depend on the rotation law through $\mathcal{H}$ multiplied by the two
weight functions $(1-\mathcal{F}_<)/k$ and $\mathcal{F}_>/k$. These functions
are both positive and smaller than one over the whole interval of integration
(see Fig.~\ref{fig:weight_functions} in the Appendix). Hence if $\mathcal{H}$
does not change its sign, \ie $\Omega$ is monotone, both integrals may at least
partly cancel each other.

The simplest monotone function one could think of as reasonable rotation law is
a power law $\Omega\propto r^\kappa$. In this case one easily computes for
(\ref{eqn:Hvphi})
\begin{equation*}
  \mathcal{H}_\kappa(r,s)=2(\kappa+1)\biggl(\frac{s}{r}\biggr)^{2(\kappa+1)}.
\end{equation*}
If we insert this into Eq.~(\ref{eqn:poisson_solution2}) the explicit dependence
on $r$ is removed from the integrals. We can evaluate them numerically if we
specify the power law exponent $\kappa$ of the rotation law. Hence the whole term
within the curly brackets depends only on $\kappa$. This implies that
$M\propto r^{2\kappa+3}$ and because $M(r)$ must be a monotonically increasing
function of radius $\kappa\geq-\frac{3}{2}$ is required. 

\begin{figure}
  \centering
  \includegraphics[width=\linewidth]{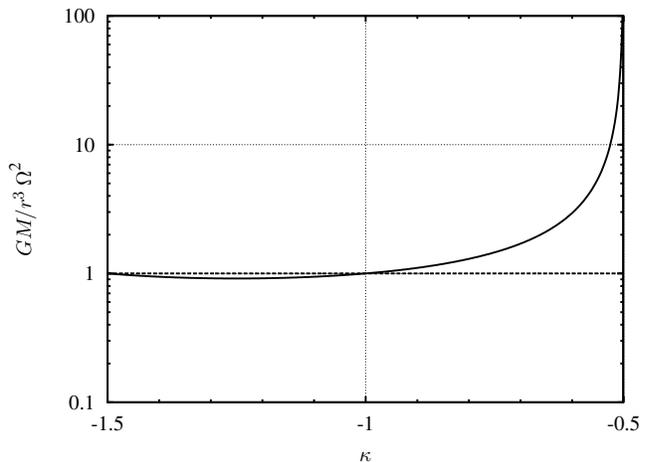}
  \caption{Deviation from monopole approximation for power law rotation curve
           $\Omega\propto r^\kappa$ as a function of the exponent.}
  \label{fig:monoerr}
\end{figure}
In Fig.~\ref{fig:monoerr} the gravitational acceleration due to the monopole term
divided by the centrifugal acceleration is shown as a function of $\kappa$. There
are two cases where the monopole approximation is exact: $\kappa=-\frac{3}{2}$
and $\kappa=-1$. The former corresponds to Keplerian rotation whereas the latter
is known as \emph{Mestel's disc} \citep[see][]{mestel1963}. The error introduced
by the monopole approximation would be less than $10\,\%$ for
$-\frac{3}{2}<\kappa<-1$ and yet for $\kappa=-\frac{3}{4}$ one overestimates the
gravitational acceleration acting inwards by less than a factor of $2$. \alert{This
observation supports the remark in \cite{lin1990} that the monopole approximation
is usually good to the $5\,\%$ level.} However, as $\kappa$ approaches $-\inv{2}$
the second integral in Eq.~(\ref{eqn:poisson_solution2}) diverges and the
monopole approximation breaks down.

\subsection{Supersonic rotation and self-gravity}
\label{sec:supersonic_rotation}

In Section~\ref{sec:slow_accretion_limit} we made an important assumption for
our model, namely that the azimuthal flow in the disc is highly supersonic
($\vphi \gg \cs$). We showed that this is a consequence of the thin disc
assumption if self-gravity is negligible. Unfortunately it is not possible to
generalize these considerations and simply apply them to self-gravitating discs,
because $\vphi$ couples to the surface density which is not known a priori.

However, if the radial balance law (\ref{eqn:monopole_approximation}) holds, we
know something about the relation between matter distribution and rotation law.
Hence we can at least check, if this relation is consistent with the assumption of
supersonic rotation. The differentiation of Eq.~(\ref{eqn:monopole_approximation})
with respect to $r$ implies that
\begin{equation}
  \label{eqn:enclosed_mass_gradient1}
  \del{M}{r} = \frac{\vphi^2}{G}\left(2x+3\right)
\end{equation}
where $x$ is the logarithmic gradient of the rotation law (Eq.~\ref{eqn:power_law_exponent}).
With the definition of the enclosed mass in Eq.~(\ref{eqn:enclosed_mass}) we can
derive another expression of its radial gradient:
\begin{equation}
  \label{eqn:enclosed_mass_gradient}
  \del{M}{r} = 2\pi\,r\sden.
\end{equation}
Hence, by equating the right hand side of Eqs.~(\ref{eqn:enclosed_mass_gradient1})
and~(\ref{eqn:enclosed_mass_gradient}) we obtain:
\begin{equation}
  \label{eqn:sigma_omega_relation}
  2\pi G \sden = r\Omega^2 \left(2x+3\right).
\end{equation}
One can use this result together with Eq.~(\ref{eqn:surface_density_integrated2})
to eliminate $\cent{\den}$ from Eq.~(\ref{eqn:scale_height_balance}). If one
furthermore substitutes point mass potential for external potential this finally
becomes
\begin{equation}
  \label{eqn:supersonic_rotation_fsg}
  \left(\frac{\cs}{\vphi}\right)^2 = \inv{\lambda}\left(2x+3\right)\frac{h}{r}
    -2(x+1)\left(\frac{h}{r}\right)^2
\end{equation}
where $\lambda$ is of order unity for reasonable values of the polytropic index
$n$ (see Fig.~\ref{fig:lambda_solution}).
This expression is the generalization of Eq.~(\ref{eqn:supersonic_rotation_nsg})
for geometrically thin self-gravitating discs in radial balance with gravitational
acceleration given by the monopole approximation. It obviously simplifies to the
non self-gravitating case in the limit $x\to-3/2$. However, if the disc 
is self-gravitating, the logarithmic derivative of $\Omega$ becomes larger than
$-3/2$ and the term of order $h/r$ on the right hand side dominates and one gets 
$(\cs/\vphi)^2 \propto h/r$. But none the less the relation $(\cs/\vphi)^2 \ll 1$
still holds and hence gravitational balance and monopole approximation are at
least consistent with the assumption of supersonic rotation in case of
geometrically thin self-gravitating discs.

\subsection{The disc evolution equation}
\label{sec:disc_equation}

In this section we will summarize the results obtained so far and derive the
disc evolution equation. The basic equations of thin disc evolution were given
in the introductory paragraph of Sec.~\ref{sec:disc_model}. In the successive 
subsections we showed that one can replace the radial momentum transport equation
(\ref{eqn:radial_momentum}) by Eq.~(\ref{eqn:monopole_approximation}) with the
enclosed mass $M(r)$ defined by Eq.~(\ref{eqn:enclosed_mass}). Because of
this it is convenient to replace the surface density in the continuity equation
(\ref{eqn:continuity}) by the enclosed mass as well and treat it as a
time-dependent function too. Hence the radial integration of (\ref{eqn:continuity})
yields
\begin{equation}
  \label{eqn:enclosed_mass_temporal_change}
  \del{M}{t} = -2\pi\,r\sden\vs.
\end{equation}
With help of Eq.~(\ref{eqn:enclosed_mass_gradient}) we may eliminate $\sden$
\citep{weizsaecker1948}:
\begin{equation}
  \label{eqn:enclosed_mass_transport}
  \del{M}{t} + \vs\del{M}{r} = 0.
\end{equation}
This replaces the continuity equation (\ref{eqn:continuity}) in the set of
model equations.

We proceed by transforming the angular momentum transport equation
(\ref{eqn:angular_momentum}). Because of Eq. (\ref{eqn:enclosed_mass_transport})
we can eliminate $\vs$ with the quotient of temporal and radial derivatives of
$M$ and with help of the balance law (\ref{eqn:monopole_approximation}) we may
replace $M$ by $r\vphi^2/G$ in these derivatives. Thus the left hand side of
Eq.~(\ref{eqn:angular_momentum}) becomes
\begin{equation*}
  \del{\sam}{t}+\vs\del{\sam}{r}
  = -\biggl(\del{\ln M}{\ln r}\biggr)^{-1} \del{\sam}{t}
  = -\frac{M}{2\pi \sden}\del{\Omega}{t}.
\end{equation*}
In the last step we used Eq.~(\ref{eqn:enclosed_mass_gradient}) and 
$\sam=r^2\Omega$. If we multiply the angular momentum equation
(\ref{eqn:angular_momentum}) by $2\pi r\sden$ and
use the result above on the left hand side it transforms to
\begin{equation*}
  -r M\del{\Omega}{t}
  =\del{}{r}\Bigl(2\pi r^2 T_{r\phi}\Bigr)
  =\del{}{r}\Bigl(\nu 2\pi r\sden r^2\partial_r\Omega\Bigr)
\end{equation*}
with the viscous stress tensor component given in (\ref{eqn:stress_tensor}).
Again we can replace $M$ by utilizing Eq.~(\ref{eqn:monopole_approximation}) and
$\sden$ with help of Eq.~(\ref{eqn:sigma_omega_relation}). Thus the final result
for the disc evolution equation becomes
\begin{equation}
  \label{eqn:disc_evolution}
  -r^4\Omega^2\del{\Omega}{t} = \del{}{r}\Bigl(
       \nu\,r^3\Omega^3 x \bigl(2x+3\bigr)\Bigr).
\end{equation}
Equation (\ref{eqn:disc_evolution}) together with the local power law exponent
$x$ defined in (\ref{eqn:power_law_exponent}) is a non-linear second order partial
differential equation which describes the advection and diffusion of angular
velocity under the influence of self-gravity and viscous friction.

If the kinematic viscosity $\nu(r,t)$ is given as a function of radial distance and
time, one may in principle derive a solution to the disc equation provided that 
one specifies appropriate initial and boundary conditions. Even if $\nu$ is not
given explicitly, the equation remains solvable if the viscosity depends on
$\Omega$ directly or indirectly through $\sden$ or $M$. We will discuss some
possibilities for an appropriate viscosity prescription applicable to accretion
discs in the next section.

The disc evolution equation derived by \cite{trefftz1952} seems to be quite
similar compared to our equation. However, we would like to emphasize that there
exists a very important difference. In contrast to our approach \cite{trefftz1952}
solves Poisson's equation for the disc potential in a pure two-dimensional
world. Therefore she obtains $M\propto\vphi^2$ as the radial balance law. This
result differs considerably from the solution we yield for an infinitesimally
thin disc embedded in a three-dimensional space.

\subsection{Viscosity prescription}
\label{sec:viscosity_prescription}

A proper description of the viscosity coefficient $\nu$ is a perennial problem
when modelling accretion discs. Although there has been a long lasting debate since
the early works of \cite{weizsaecker1948} and \cite{luest1952} on the nature of
the viscosity coefficient it became a broad consensus that molecular viscosity is
not sufficient to explain the accretion process \citep{frank2002}.
\alert{In case of non self-gravitating discs}
there are basically two processes that lead to sufficiently high stresses:
Strong, large scale magnetic fields and turbulence \citep{shakura1973}.

\alert{However, in self-gravitating discs there exists a completely different
mechanism for redistribution of angular momentum. If these discs become
gravitationally unstable, they can transfer angular momentum via compressible
density waves and dissipate energy in shocks \citep{balbus1999}. Since the amount
of angular momentum transfer depends on the details of the inhomogeneous structures,
one usually has to perform multidimensional simulations to investigate these
processes. Thus it has been the fundamental question of self-gravitating
accretion disc theory in the past 20 years whether or not it is
possible to model those discs with a simple one-dimensional diffusion model.
Since we do not want to repeat ourselves we refer to Sec.~\ref{sec:introduction}
for a short review of the most important findings including important references.
We proceed summarizing the discussion on this issue with the statement that the
viscous diffusion approximation seems possible even for self-gravitating discs
if they are geometrically thin and not too massive.}

\alert{In the context of non self-gravitating discs a very popular description of the
effective shear stresses} was proposed by \cite{shakura1973} who derived a
parametrization which couples the dominant stress tensor component $t_{r\varphi}$
to pressure
\begin{equation}
  \label{eqn:alpha_model}
  t_{r\varphi} = -\alpha P
\end{equation}
with model parameter $0<\alpha < 1$. If we carry out vertical integration according
to Eqs.~(\ref{eqn:surface_pressure}) and (\ref{eqn:stress_tensor}), we can cast
this equation with help of Eq.~(\ref{eqn:vertical_structure_solution}) into an
expression for the effective turbulent viscosity \cite[see][]{lin1990}
\footnote{In case of negligible self-gravity one can utilize
Eq.~(\ref{eqn:supersonic_rotation_nsg}) to transform the viscosity prescription
into the well-known formula $\nu=\tilde{\alpha} h\cs$.}
\begin{equation}
  \label{eqn:alpha_viscosity_general}
  \nu = \tilde{\alpha} \frac{\cs^2}{\Omega}.
\end{equation}
The new parameter $\tilde{\alpha}=-\alpha\eta/x$ is again smaller than one,
because $\eta\leq 1$ (see Eq.~(\ref{eqn:definition_eta})) and the logarithmic
derivative of $\Omega$ (Eq.~\ref{eqn:power_law_exponent}) is negative and of
order one. \cite{balbus1991} have discussed an instability that in magnetic
accretion discs can give rise to viscous stresses of the $\alpha$ type with --
if only marginally -- the required strength.

The $\alpha$ viscosity couples to the midplane speed of sound $\cs$ which in
turn depends on midplane temperature. Therefore it is generally not possible
to compute the viscosity coefficient without knowing anything about the
temperature distribution within the disc. Unfortunately this would require
solving the energy equation in addition to the disc evolution equation derived
in the previous section. \cite{paczynski1978} was the first who came up with the
idea of self-regulation \cite[see also][]{bertin1997,bertin1999,lodato2007} which circumvents
the solution of the energy equation by simply proposing that the flow in
self-gravitating discs is always at the border of instability $Q\approx 1$ with
the Toomre parameter \cite[see][]{toomre1964}
\begin{equation}
  \label{eqn:toomre_paramter}
  Q = \frac{\cs \alert{\kappa_\mathrm{e}}}{\pi G\sden}\approx\frac{\cs \Omega}{\pi G\sden}
\end{equation}
where $\alert{\kappa_\mathrm{e}}$ is the epicyclic frequency
\begin{equation*}
  \alert{\kappa_\mathrm{e}}=\Omega\sqrt{\del{\ln \sam^2}{\ln r}}=\Omega\sqrt{2x+4}\approx\Omega.
\end{equation*}
The error induced by the approximation is of order unity for any reasonable value
of $x$. If we solve Eq.~(\ref{eqn:toomre_paramter}) for $\sden$ and insert the
result into Eq.~(\ref{eqn:sigma_omega_relation}) we obtain an expression for
$\cs$ in terms of $\vphi$ and $Q$:
\begin{equation}
  \label{eqn:self_regulation}
  \cs = \left(x+\tfrac{3}{2}\right)Q\vphi
\end{equation}
which allows us to eliminate $\cs$ from the viscosity prescription
(\ref{eqn:alpha_viscosity_general}). Hence we finally get
\begin{equation}
  \label{eqn:beta_viscosity_LP}
  \nu = \tilde{\beta} (2x+3)^2 r^2\Omega
\end{equation}
with the new parameter $\tilde{\beta}=\tilde{\alpha}Q^2/4$. We would like to mention that
the assumption of marginal stability, \ie $Q\approx 1$,  might be a problem, 
because this would contradict the assumption of supersonic motion for
self-gravitating discs where the logarithmic derivative of the rotation law
deviates from its Keplerian value $x=-3/2$ (see Eq.~(\ref{eqn:self_regulation})
and discussion in Sec.~\ref{sec:supersonic_rotation}). The viscosity
prescription in Eq.~(\ref{eqn:beta_viscosity_LP}) was first proposed by
\citet[hereafter LP]{lin1987} who assumed that the turbulent viscosity is caused
by some kind of gravitational instability:
\begin{equation*}
  \nu = L_\mathrm{crit}^2 \Omega = \biggl(\frac{G\sden}{\Omega^2}\biggr)^2 \Omega
\end{equation*}
where $L_\mathrm{crit}$ is the typical length scale of unstable modes. If we
eliminate $\sden$ using Eq.~(\ref{eqn:sigma_omega_relation}) we can transform
this equation into
\begin{equation*}
  \nu = \biggl(\frac{2x+3}{2\pi}\biggr)^2 r^2\Omega
\end{equation*}
which becomes exactly (\ref{eqn:beta_viscosity_LP}) if we set $\tilde{\beta}=1/4\pi^2\approx 0.025$.
It was already mentioned by \cite{bertin1999} and others \cite[see][]{lin1990}
that LP viscosity prescription fails inevitably in the Keplerian limit, because
self-regulation can only occur in the self-gravitating regime where the disc can
become gravitationally unstable. The consequence is that $\nu$ vanishes as
$x\to-3/2$. The reason for this odd behaviour is the assumption that one can
assign a finite and fixed value to $Q$ and therefore $\tilde{\beta}$. However, in the
Keplerian limit the disc must be Toomre stable and therefore $Q$ must become
much larger the one. This could compensate the factor $(2x+3)^2$ in the
viscosity prescription which tends to zero as $x\to-3/2$. To shed more light on
this, we insert the ratio $\cs/\vphi$ from Eq.~(\ref{eqn:self_regulation})
into Eq.~(\ref{eqn:supersonic_rotation_fsg}). If we assume that the ratio $h/r$
is small, but always larger than zero in the Keplerian limit as $x\to-3/2$, we
conclude that the left hand side of
\begin{equation*}
  \left(x+\tfrac{3}{2}\right)^2 Q^2 = \left(2x+3\right)\frac{h}{r}
    -2(x+1)\left(\frac{h}{r}\right)^2
\end{equation*}
must remain greater than zero too and therefore the viscosity should not vanish if
self-gravity becomes negligible. However it might become rather small, because
$h/r\ll 1$ in thin discs.

The simplest modification of (\ref{eqn:beta_viscosity_LP}) to overcome
these limitations would be to assume that $\nu\propto r^2\Omega$. A viscosity
of this kind was proposed by \citet[\citeyear{duschl2000}, hereafter DSB]{duschl1998}
who assumed that the effective Reynolds number does not fall below the critical
Reynolds number. The so-called $\beta$-ansatz just reads
\begin{equation}
  \label{eqn:beta_viscosity_DSB}
  \nu = \beta\,r\vphi = \beta\,r^2\Omega.
\end{equation}
where the constant parameter $\beta$ is given by the inverse of the critical
Reynolds number. 
\begin{alertenv}
  This is still a parameterization and not an ab-initio solution to the problem.
  The ansatz (\ref{eqn:beta_viscosity_DSB}), however, has several promising
  aspects:
  \begin{enumerate}
    \item For \emph{fully self-gravitating disk} regions, which are dominated by
      the mass distribution of the disk in the direction perpendicular to the
      rotational plane as well as in the rotational plane itself, ansatz
      (\ref{eqn:beta_viscosity_DSB}) is equivalent to the hypothesis that the
      ratio of dynamic timescale ($\tau_\mathrm{dyn}\sim\Omega^{-1}$) and viscous
      timescale ($\tau_\mathrm{vis}\sim r^2/\nu$) is constant. In other words
      one assumes a linear relation between the two relevant timescales:
      $\tau_\mathrm{vis}=\beta^{-1}\tau_\mathrm{dyn}$. \cite{duschl2006} succeed
      in showing numerically that instabilities in a self-gravitating flow lead
      to a viscosity of the functional form of the $\beta$-ansatz, indeed.
      This is furthermore supported by the observation that in the fully
      self-gravitating limit, where the rotation law deviates considerably
      from Keplerian motion ($x>-3/2$), the $\beta$-prescription approaches the
      values obtained with the self-regulated $\alpha$-ansatz.
      (\ref{eqn:beta_viscosity_LP}). 
    \item For \emph{Keperian self-gravitating disk} regions, the vertical
      structure is dominated by local self-gravity, \ie, the local mass
      distribution in the disk, while in the rotational plane the (almost)
      equilibrium between gravitational and centrifugal forces is still
      determined by the central mass, \ie, the rotational velocity is Keplerian.
      There a functional form of (\ref{eqn:beta_viscosity_DSB}) solves the
      problem of a quasi-thermostat as discussed by \cite{duschl2000}.
    \item In the limit of negligible self-gravity (\emph{Keplerian disk} regions)
      it smoothly merges into the $\alpha$-prescription \citep{duschl1998,duschl2000}.
      This can be seen by plugging the non-self-gravitating (Keplerian) scale-height
    (\ref{eqn:scale_height_kepler}) in the $\beta$-viscosity prescription
    (\ref{eqn:beta_viscosity_DSB}):
    \begin{equation*}
      \nu = \beta \frac{\vphi^2}{\Omega}=\beta\left(\frac{r}{h}\right)^2\frac{\cs^2}{\Omega}.
    \end{equation*}
    A comparison with the $\alpha$-ansatz (Eq.~\ref{eqn:alpha_viscosity_general})
    reveals the above mentioned relation\footnote{\alert{The aspect ratio $h/r$ in
    Keplerian discs usually depends only weakly on $r$. It is roughly of the
    order of $10^{-1}\dots10^{-2}$ in protoplanetary discs \citep{andrews2009}
    and an order of magnitude smaller in AGN discs \citep{collin1990,lin1996}.}}. Here,
    $\beta$ is equivalent to a constant Reynolds number.
  \end{enumerate}
  In that respect the $\beta$-ansatz is, at least formally, a generalization of
  the classical $\alpha$-ansatz, which otherwise runs into trouble in the fully
  and Keplerian self-gravitating domains. To couple all three regions, $\beta$ 
  needs to be chosen such that a smooth transition between them can be achieved.
  If, similar to the reasoning for classical accretion disks, one takes resort to
  time scale arguments, it turns out that the standard values of $\alpha$,
  which are of the order of $10^{-2\dots 0}$, are compatible with values of $\beta$
  which correspond to those of the inverse of the critical Reynolds
  number\footnote{\alert{Despite the fact that in a gravitationally driven disk, the
  Reynolds number loses its meaning, a number with a similar functional dependence
  shows up in the self-gravitating context, too, as the ratio of viscous and
  dynamical time scales, $\tau_\mathrm{vis}$ and $\tau_\mathrm{dyn}$,
  respectively. In that respect, assuming a constant value of $\beta$ is
  tantamount to a constant ratio of the two time scales.}}.
\end{alertenv}
Some aspects of this have also been investigated by
\citet[hereafter RZ]{richard1999} who proposed a quite similar prescription:
\begin{equation}
  \label{eqn:beta_viscosity_RZ}
  \nu = \beta\,r^3\biggl|\del{\Omega}{r}\biggr|
  = \beta\,\left|x\right|\, r^2\Omega.
\end{equation}
It differs from the previous one by a factor of $|x|$ which is of order unity
in most cases.

All the parametrizations described above differ only in their dependence on the
local power law exponent $x$. Hence we may combine the three viscosity functions
into one formula and introduce a new function $f(x)$ to distinguish between these
prescriptions
\begin{equation}
  \label{eqn:definition_combined_viscosity}
  \nu = \beta r^2 \Omega f(x)
  \quad\textnormal{with}\quad
  f(x)=\begin{cases}
         1 & \textnormal{DSB} \\
         |x| & \textnormal{RZ} \\
         (2x+3)^2 & \textnormal{LP}
       \end{cases}
\end{equation}
The magnitude of the viscous coupling constant $\beta$ proposed by LP and DSB
is of order $10^{-2}$ to $10^{-3}$ whereas RZ derived smaller values of
approximately $4\cdot 10^{-6}$. However, the precise value of $\beta$ is not of
substantial importance for the solution of the disc evolution equation. If we
define the new time variable $\tau = \beta\,t$ we can rewrite
Eq.~(\ref{eqn:disc_evolution}) thereby eliminating $\beta$ from the equation
\begin{equation}
  \label{eqn:disc_evolution_with_viscosity}
  -r^4\Omega^2\del{\Omega}{\tau} = \del{}{r}\Bigl(
       r^5\Omega^4 x (2x+3) f(x)\Bigr).
\end{equation}
Hence we can transform any solution $\Omega(r,\tau)$ back to $\Omega(r,t)$ 
by rescaling the time variable. Although Eq.~(\ref{eqn:disc_evolution_with_viscosity})
does not depend on $\beta$, its actual value may have an impact on the solution
due to boundary conditions (see Sec.~\ref{sec:solution_procedure}).

\section{Self-similar solutions}
\label{sec:self-similar_solutions}

In this section we show how to solve Eq.~(\ref{eqn:disc_evolution_with_viscosity})
using similarity methods\footnote{For an elaborate discussion of Lie groups,
similarity methods and related topics the reader may consult the textbook by
\cite{bluman2002}.}. To simplify the derivation we introduce non-dimensional
variables and functions and rewrite the disc evolution equation in terms of
these. If we specify an arbitrary length scale $\tilde{r}$ and mass scale
$\widetilde{M}$ and define the time scale $\tilde{\tau}$ according to
\begin{equation}
  \label{eqn:time_scale}
  \tilde{\tau} = \sqrt{\frac{\tilde{r}^3}{G\,\widetilde{M}}}
\end{equation}
we can non-dimensionalize all variables, functions and equations derived in
the previous sections. The equations retain their form except for those
containing Newton's gravitational constant $G$ which must be set to unity
in the non-dimensional equations. Once we have solved the self-gravitating
accretion disc problem for the non-dimensional functions we can use these
scaling parameters to switch back to the real world quantities. In
Sec.~\ref{sec:application} we discuss the self-similar evolution of massive
AGN accretion discs and demonstrate how to apply the inverse transformations to
recover the dimensioned quantities.

\subsection{Differential equation of self-similar evolution}
\label{sec:differential_equation}

A short calculation yields that Eq.~(\ref{eqn:disc_evolution_with_viscosity})
is invariant under the family of one-parameter Lie groups of scaling transformations
\begin{equation}
  \label{eqn:stretching_group}
  r'=\lambda^a r,\qquad
  \tau'=\lambda^b \tau,\qquad
  \Omega'=\lambda^c \Omega
\end{equation}
if and only if $c=-b$. The group invariants\footnote{These assignments are
ambiguous because any function of the group invariants is again a group
invariant.} are
\begin{equation}
  \label{eqn:group_invariants}
  \xi=r\tau^{1/\kappa}\quad\textnormal{and}\quad y=-(\kappa\Omega\tau)^{-1}
\end{equation}
with $\kappa=-b/a$ the general group invariant solution is determined by
the expression $F(y,\xi)=0$ where $F$ is a (not yet determined) function of the
group invariants. This is an implicit definition of the function $y(\xi)$. Thus
with Eq.~(\ref{eqn:group_invariants}) one can write down the explicit form of
the group invariant solutions
\begin{equation}
  \label{eqn:group_invariant_solution}
  \Omega(r,\tau) = -\inv{\kappa\tau\,y\bigl(\xi(r,\tau)\bigr)}.
\end{equation}
The remaining problem is to determine $y$ as a function of the similarity
variable $\xi$.

If we apply the original PDE (\ref{eqn:disc_evolution_with_viscosity}) with $x$
given by Eq.~(\ref{eqn:power_law_exponent}) to the group invariant solutions we
obtain a coupled system of ordinary differential equations (ODEs) for $x(\xi)$
and $y(\xi)$
\begin{align}
  \label{eqn:selfsim_ode1}
  \dd{x}{\ln\xi} &= \frac{(x-\kappa)y - (4x+5)z(x)}{z'(x)} \\
  \label{eqn:selfsim_ode2}
  \dd{y}{\ln\xi} &= -xy
\end{align}
where $z'(x)$ is the derivative with respect to $x$ of the viscosity dependent
function
\begin{equation}
  \label{eqn:definition_zx}
  z(x)=x(2x+3)f(x)
  =\begin{cases}
         x(2x+3) & \textnormal{DSB} \\
         -x^2(2x+3) & \textnormal{RZ} \\
         x(2x+3)^3 & \textnormal{LP}
   \end{cases}
\end{equation}
and $f(x)$ is given by Eq.~(\ref{eqn:definition_combined_viscosity}). In case of
the RZ-prescription we assumed that $x<0$, \ie $\Omega$ always decreases as
$r\to\infty$. The remaining constant $\kappa$ in Eq.~(\ref{eqn:selfsim_ode1})
is a free parameter which has to be determined from the auxiliary conditions (see
Section~\ref{sec:auxiliary_conditions}).

The planar differential system (\ref{eqn:selfsim_ode1},\ref{eqn:selfsim_ode2})
is an autonomous set of two first order ODEs. It is always possible to reduce
these systems by elimination of the independent variable to a single first order
equation
\begin{equation}
  \label{eqn:selfsim_ode3}
   \dd{x}{y} = \frac{(4x+5)z(x)-(x-\kappa)y}{x z'(x) y}.
\end{equation}
The solution of Eq.~(\ref{eqn:selfsim_ode3}) gives $x(y)$ which can be used to
integrate Eq.~(\ref{eqn:selfsim_ode2})
\begin{equation}
  \label{eqn:selfsim_first_integral}
  \ln\xi = -\int\frac{\di{y}}{x(y)y}.
\end{equation}
This yields $\xi(y)$ and its inverse $y(\xi)$ which is required to compute the
self-similar solution $\Omega(r,\tau)$  given in Eq.~(\ref{eqn:group_invariant_solution}).
Unfortunately, Eq.~(\ref{eqn:selfsim_ode3}) is a non-linear ODE and one cannot
expect to solve it in terms of known functions, in general. In fact, the
inverse of Eq.~(\ref{eqn:selfsim_ode3}) is an \emph{Abel Equation} of the
second kind. We were not able to identify a non-linear transformation which maps
our equation to any of the few classes with known analytical solutions
\cite[see, \eg][]{polyanin2003}.

Although analytical solutions are not excluded per se, we focused on numerical
methods to solve the self-similar disc equations. The task is basically to
solve the autonomous system of ODEs (\ref{eqn:selfsim_ode1},\ref{eqn:selfsim_ode2})
which can be achieved using standard techniques. However, the system under
investigation exhibits some pitfalls such as singular points where the solution
is not unique. Careful treatment of these peculiarities is essential. We
discuss the numerical solution procedure in detail in
Sec.~\ref{sec:numerical_method}.

\subsection{Auxiliary conditions}
\label{sec:auxiliary_conditions}

The problem addressed in the previous sections is a so-called initial-boundary-value
problem. This problem is well defined only if we provide auxiliary conditions
in addition to the PDE (\ref{eqn:disc_evolution_with_viscosity})
describing the dynamical evolution of the function $\Omega(r,\tau)$. Thus we
have to define some initial state $\Omega_0(r)$ at time $\tau=0$ from which
the solution evolves and we must specify how $\Omega$ changes on the boundary
of the spatial domain for all $\tau>0$. Since the spatial domain we are
interested in is the positive real axis, its boundary is determined by the
limits $r=0$ and $r\to\infty$. Therefore one has to specify three auxiliary
conditions.

If we are looking for self-similar solutions, we demand an additional restriction
to the set of possible solutions which must be related to the auxiliary
conditions. The self-similar solutions are obtained from the integration of the
first order system of ODEs (\ref{eqn:selfsim_ode1}, \ref{eqn:selfsim_ode2}). In
contrast to the original PDE, this problem is well
posed if we supply initial conditions for $x$ and $y$. Hence the number of
necessary auxiliary conditions is reduced by one. As a consequence self-similar
solutions only exist if two of the original conditions coalesce
\cite[see][Chap.~4.3]{ames1965}. Thus, although one can normally choose
arbitrary auxiliary conditions, the requirement of self-similarity imposes some
restrictions (see Sec.~\ref{sec:boundary_conditions}).

In addition, one has to ensure that these conditions are consistent with
the underlying physical model. In case of the disc evolution model described above
there are some constraints due to the fact that the surface density $\sden$ must
be positive everywhere at any time. The same applies to the enclosed mass which
is not only positive but also a monotonically increasing function of radial
distance $\partial_r M \geq 0$ for all $r>0$ (see Eq.~(\ref{eqn:enclosed_mass_gradient})).
Since $\Omega$ is related to $M$ via Eq.~(\ref{eqn:monopole_approximation})
the auxiliary conditions (and the solutions too) must fulfill
\begin{equation}
  \label{eqn:omega_constraint1}
  r^3\Omega^2 > 0
  \quad\Leftrightarrow\quad
  \Omega\ne 0
\end{equation}
and
\begin{equation}
  \label{eqn:x_constraint}
  \del{\ln r^3\Omega^2}{\ln r} = 2x+3 \geq 0
  \quad\Leftrightarrow\quad
  x \geq -\tfrac{3}{2}
\end{equation}
for all $0<r<\infty$ and $0\le \tau<\infty$. Condition~(\ref{eqn:omega_constraint1})
may become even more restrictive if we demand that $\Omega$ must be continuous.
Therefore we conclude that
\begin{equation}
  \label{eqn:omega_constraint2}
  \Omega > 0.
\end{equation}
The sign is determined by the orientation of the rotational axis, which we
define to point always in the positive $z$-direction. 

\subsubsection{Initial conditions}
\label{sec:initial_conditions}

Let us suppose that the initial condition $\Omega_0(r)$ satisfies the relations
(\ref{eqn:x_constraint}, \ref{eqn:omega_constraint2}). With
Eqs.~(\ref{eqn:group_invariants}) we obtain
\begin{equation*}
  \Omega_0(r) = -\lim_{\tau\to 0}\bigl(\kappa\tau y(\xi(r,\tau))\bigr)^{-1}
  = -\kappa r^\kappa \lim_{\xi\to\infty}\bigl(\xi^\kappa y(\xi)\bigr)^{-1}
\end{equation*}
where we assumed that $\kappa<0$ (which will be justified below). The limit on
the right hand side is independent of $r$ and should be finite. Hence we
conclude that
\begin{equation}
   \label{eqn:initial_condition_nondim}
   y(\xi) \propto \xi^{-\kappa} \quad\mathrm{as}\quad \xi\to\infty
\end{equation}
which has the implication that the initial condition for self-similar solutions
must be a power law of radius
\begin{equation}
  \label{eqn:initial_condition}
  \Omega_0(r) \propto r^\kappa.
\end{equation}
In view of this result the assumption that $\kappa<0$ seems reasonable because
otherwise the initial $\Omega_0$ would be an increasing function of radius
which we want to exclude from our considerations (see
Sec.~\ref{sec:monopole_approximation}). We therefore conclude that the so far
unknown parameter $\kappa$ in Eq.~(\ref{eqn:selfsim_ode1}) introduced by the
requirement of group invariance imposed on the solution is simply the power
law exponent of the initial condition.

We already discussed in Section~\ref{sec:monopole_approximation} that rotation
laws $\Omega \propto r^\kappa$ with an exponent $\kappa$ greater than
$-\tinv{2}$ cause infinite centrifugal forces as $r\to\infty$. Furthermore we
showed that the monopole approximation breaks down if the power law exponent
approaches $-\tinv{2}$. Taking into account that Eq.~(\ref{eqn:x_constraint})
holds, one should therefore demand that
\begin{equation}
  \label{eqn:kappa_constraint}
  -\tfrac{3}{2}\leq \kappa\lesssim-\tfrac{3}{4}.
\end{equation}
This restricts the parameter $\kappa$ to a very limited range of accessible
values between centrally condensed mass distributions and configurations with 
flattened rotation curve where the mass is initially dispersed over the whole
disc.

\subsubsection{Boundary conditions}
\label{sec:boundary_conditions}

Normally one can think of a variety of physically meaningful boundary conditions
for the self-gravitating accretion disc problem. However, in case of self-similar
solutions the selection of valid boundary conditions is restricted, as was
already mentioned above. Since $\xi$ depends on $r$ and $\tau$ according to
(\ref{eqn:group_invariants}) we have
\begin{equation*}
  \xi\to\infty \Leftrightarrow
  \begin{cases}
    \tau\to 0 &\textnormal{for any fixed}\quad 0<r<\infty\\
    r\to\infty &\textnormal{for any fixed}\quad 0<\tau<\infty
  \end{cases}
\end{equation*}
if $\kappa<0$. Therefore the outer boundary condition must coalesce with the
initial condition.

This result becomes quite clear, if one remembers that the dynamic time scale
for the discs evolution is roughly given by the inverse of $\Omega$. Since the
initial condition is a decreasing power law, its inverse tends to infinity as
$r$ approaches infinity. The evolutionary time scale becomes infinitely large
which means that the disc does not evolve at all and stays in its initial state
at the outer rim.

At the inner boundary there are basically two types of reasonable boundary
conditions: Vanishing and finite torque supplied at the origin. The viscous
torque $G(r,\tau)$ is given by
\begin{equation}
  \label{eqn:viscous_torque}
  G(r,\tau) = 2\pi r^2 \Trphi
\end{equation}
where the stress tensor component $\Trphi$ is defined in Eq.~(\ref{eqn:stress_tensor}).
One easily verifies that this could be rewritten in terms of $x(\xi)$, $y(\xi)$
and the similarity variable $\xi$:
\begin{equation}
  \label{eqn:viscous_torque_nondim}
  G(r,\tau)=\beta\kappa^{-4}\,\tau^{-(5/\kappa+4)}~\xi^5\,y^{-4}\,z(x).
\end{equation}
For any fixed and finite time $\tau$ Eq.~(\ref{eqn:group_invariants}) yields
\begin{equation*}
  \xi \to 0 \Leftrightarrow r \to 0
\end{equation*}
We therefore conclude that the torque vanishes at the inner boundary at any time
$0<\tau<\infty$ if
\begin{equation}
  \label{eqn:bc_vanishing_torque}
  \lim\limits_{\xi\to0} \xi^5\,y^{-4}\,z(x) = 0.
\end{equation}
An alternative would be that this limit is finite. In that case there are
three distinct ways how the torque acts on the inner boundary depending on the
value of $\kappa$, namely
\begin{align*}
  &\kappa<-\tfrac{5}{4} \qquad\textnormal{the torque decreases with time} \\
  &\kappa=-\tfrac{5}{4} \qquad\textnormal{the torque remains constant} \\
  &\kappa>-\tfrac{5}{4} \qquad\textnormal{the torque increases with time.}
\end{align*}
Therefore a constant torque boundary condition is only applicable if the initial
condition has a unique slope.
\begin{alertenv}
  We would like to emphasize that this restriction is due to the requirement
  of self-similarity imposed on the solutions and not a limitation of our disc
  model. However, one should not be worried too much, because in any realistic
  accretion disc scenario one would not expect that the torque acting at the
  inner boundary remains constant over a significant time span. In fact the
  most reasonable assumption in case of discs around black holes seems to be
  the zero torque boundary condition, because the event horizon prohibits any
  coupling to the central object. Another possibility would be a central object
  of finite size like an inner -- possibly geometrically thick -- accretion disc
  or a spinning protostar with spatial extent much smaller than the typical
  scales of the surrounding self-gravitating disc. In that case the torque acting
  at the inner boundary may indeed be finite and it is most likely that it
  decreases with time as the central object spins down.
\end{alertenv}

\subsubsection{Conservation conditions}
\label{sec:conservation_conditions}

Instead of the initial condition one could also specify the required auxiliary
condition using a different but equivalent formulation. Let's define
\begin{equation*}
  \mathcal{C}(\tau) = 2\pi \int_0^{R(\tau)} \sden \Omega^a r^b r \di{r}
\end{equation*}
with arbitrary constant exponents $a$ and $b$. The quantity $\mathcal{C}$ is
conserved if
\begin{equation}
  \label{eqn:conservation_condition}
  \dd{\mathcal{C}}{\tau} = 0.
\end{equation}
For example if $a=0$ and $b=0$ then $\mathcal{C}$ is the enclosed mass within
radial distance 
\begin{equation*}
  R(\tau) = \xi_0\tau^{-1/\kappa}
\end{equation*}
where $\xi_0$ is a fixed value of the similarity variable $\xi$
(Eq.~\ref{eqn:group_invariants}). We may rewrite the definition of $\mathcal{C}$
using non-dimensional variables and functions:
\begin{equation}
  \label{eqn:conserved_quantity}
  \mathcal{C}(\tau) = \frac{\int_0^{\xi_0} (2x+3) y^{-(a+2)} \xi^{b+2} \di{\xi}}%
    {\kappa^{(a+2)}\tau^{(a+2)+(b+3)/\kappa}}
\end{equation}
Since $\mathcal{C}$ must be finite in order to be conserved, the integral in the
nominator should give a finite number and condition (\ref{eqn:conservation_condition})
is fulfilled if the exponent of $\tau$ in the denominator vanishes, \ie
\begin{equation}
  \label{eqn:conservation_condition_A}
  \kappa = -\frac{b+3}{a+2}.
\end{equation}
The initial power law exponent $\kappa$ is completely determined by the numbers
$a$ and $b$ which define the conserved quantity $\mathcal{C}$. Therefore one
concludes that each initial condition generates a self-similar solution with
specific conservation properties.

For example, if $a=0$ and $b=0$, $\mathcal{C}$ is the enclosed disc mass within
radius $R(\tau)$ which is conserved if $\kappa=-\frac{3}{2}$. This is a
reasonable result because in that case the mass of the disc must be finite and
therefore the rotation law at the outer boundary as $r\to\infty$ is necessarily
Keplerian. We will discuss  these solutions in more detail in Sec.~\ref{sec:time_evolution}.

Another interesting case is $a=1$ and $b=2$ where $\mathcal{C}$ corresponds to
the discs angular momentum. Unfortunately, this leads to $\kappa=-\tfrac{5}{3}$
which conflicts with the requirement in Eq.~(\ref{eqn:kappa_constraint}). Thus a
physically meaningful self-similar solution with conserved angular momentum
within the disc throughout the whole evolution does not exist. This does not
mean that all other solutions violate angular momentum conservation. Instead,
these discs do exactly what they are supposed to do, namely they transfer
angular momentum from the inner regions to the environment beyond the radius
$R(\tau)$.

\subsection{Numerical method}
\label{sec:numerical_method}

As already mentioned in Section~\ref{sec:differential_equation} the initial
value problem defined by the system of non-linear ODEs~(Eqs.~\ref{eqn:selfsim_ode1}
and~\ref{eqn:selfsim_ode2}) can be solved numerically with standard techniques.
We use the programs \texttt{ode} \citep*{tufillaro1992} and \texttt{lsode}
\citep{hindmarsh1983} which implement different numerical integration schemes
including explicit and semi-implicit methods such as Runge-Kutta-Fehlberg and
predictor-corrector Adams-Moulton schemes as well as fully implicit schemes like
Gear's method which utilizes Backward Differentiation Formulas (BDF)
\citep*[see][and references therein]{hairer1993}. 

For comparison we tried another numerical scheme which solves differential
algebraic equations (DAE). Thereby we must transform the system of ODEs
into a system of DAEs which was simply done by adding $z(\xi)$ to the set of
independent functions. Then rewriting Eq.~(\ref{eqn:selfsim_ode1}) yields a
differential equation for $z(\xi)$ which must be combined with
Eq.~(\ref{eqn:selfsim_ode2}) and the algebraic constraint for $x(\xi)$ given by
Eq.~(\ref{eqn:definition_zx}). As for ODEs there is a variety of different
numerical methods available to solve DAEs \citep[see][]{hairer1996,petzold1998}.
We used the program \texttt{daspk} described in \cite*{brown1994}.

We found that the results obtained with the different programs and numerical
schemes are similar within limits of numerical errors. In view of performance
and numerical efficiency we would certainly recommend the BDF scheme and the DAE
solver which allow for larger step sizes. This is important for large $\xi$
because the system becomes stiff in this limit.

Moreover, in order to solve the problem one has to determine feasible initial
values for the numerical integration. This basically implies that the solutions
obtained with these initial values must satisfy the restrictions and the
auxiliary conditions discussed in the previous section. But there are still some
free parameters, namely the proportionality constant in the initial condition
(Eq.~\ref{eqn:initial_condition}) and the torque applied to the disc at the
inner boundary (Eq.~\ref{eqn:viscous_torque_nondim}).

\subsubsection{Phase Plane}
\label{sec:phase_plane}

\begin{figure}
  \centering
  \includegraphics[width=\linewidth]{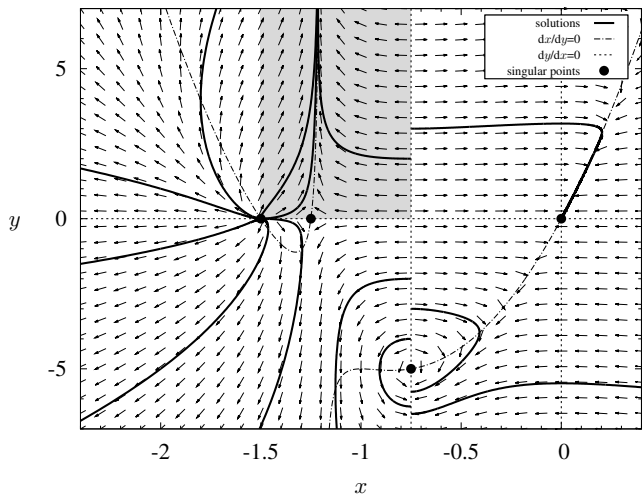}
  \caption{Phase diagram of the planar differential system defined by
     Eqs.~(\ref{eqn:selfsim_ode1},\ref{eqn:selfsim_ode2})
     for DSB viscosity prescription with $\kappa=-6/5$. The arrows show the
     local gradient of the integral curves with direction pointing towards
     increasing values of the parameter $\xi$ on the curve. Some selected
     solutions obtained for different initial values are shown as solid
     curves.}
  \label{fig:phasediagram_big}  
\end{figure}
In this section we examine the phase plane of the planar differential system
to get an idea of the solutions topology. By doing so we can check which part of
the parameter space is admitted by the restrictions given above. Since the
differential equations depend on the parameter $\kappa$ and the viscosity
prescription through the function $z(x)$~(Eq.~\ref{eqn:definition_zx}) the
phase diagram shown in Fig.~\ref{fig:phasediagram_big} would change if these
were varied. However, the main features remain unchanged. Among these are
\begin{itemize}
  \item two nodes on the $x$-axis at $x=-\tfrac{3}{2}$ and $x=0$
  \item one saddle point on the $x$-axis at $x=-\tfrac{5}{4}$
  \item one singular vertical line.
\end{itemize}
The location of the singular line $x_\mathsf{c}$ depends on the viscosity
prescription ($-\tfrac{3}{4}$ for DSB, $-1$ for RZ and $-\tfrac{3}{8}$ for LP).
In addition, there exists another singular point -- except for the case where
$\kappa=x_\mathsf{c}$ -- which is always located on the singular line. On the
singular line the numerical integration fails, because the denominator on the
right hand side of Eq.~(\ref{eqn:selfsim_ode1}) vanishes which causes the
derivative of $x$ with respect to $\ln\xi$ to become infinitely large as
$x\to x_\mathsf{c}$. The only way a solution may continuously pass it, would be
through the fourth singular point where the numerator vanishes simultaneously.
The type of this singular point and its location on the singular line depends on
viscosity prescription and $\kappa$:
\begin{equation*}
  \textnormal{DSB:}\,\frac{9}{4\kappa+3}, \quad
  \textnormal{RZ:}\,\frac{1}{\kappa+1}, \quad
  \textnormal{LP:}\,\frac{15309/128}{8\kappa+3}.
\end{equation*}
The point disappears (its $y$-value tends to infinity) in all cases if $\kappa$
equals the location of the singular line $x_\mathsf{c}$. It lies below the
$x$-axis if $\kappa<x_\mathsf{c}$  and otherwise above it.

In Fig.~\ref{fig:phasediagram_big} the shaded area marks the region where the
solutions match the requirements~(\ref{eqn:x_constraint})
and~(\ref{eqn:omega_constraint2}). In addition we set an upper limit on $x$ to
exclude rather flat rotation curves where the monopole approximation no longer
holds (see Sec.~\ref{sec:monopole_approximation}). There are basically two
distinct families of integral curves of interest. One starts somewhere on the
singular line $x_\mathsf{c}>\kappa$ and approaches $y\to+\infty$ as
$x\to\kappa^{+}$ from above and the other originates at the singular point
$(-\tfrac{3}{2},0)$ and approaches $y\to+\infty$ as $x\to\kappa^{-}$.
Both families exhibit similar asymptotic behavior for large $y$ which is
compliant with the required initial and outer boundary conditions, but
only the second family admits solutions with the correct asymptotic behavior
as $y\to 0$ namely $x\to-\frac{3}{2}$, \ie Keplerian motion as $r\to 0$.

\begin{figure}
  \centering
  \includegraphics[width=\linewidth]{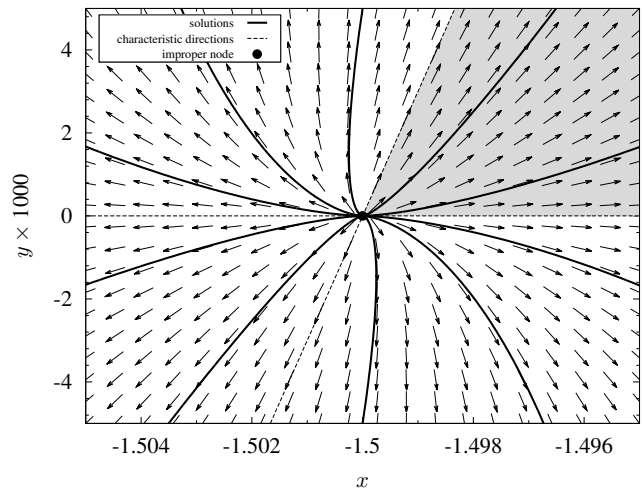}
  \caption{Phase diagram around the improper node for DSB viscosity prescription
     with $\kappa=-1$; values on the $y$-axis are scaled by a factor of $1000$.}
  \label{fig:phasediagram_node}  
\end{figure}
Further analysis therefore focuses on the shape of the integral curves in the
vicinity of the singular point at $\left(-\tfrac{3}{2},0\right)$ which
corresponds to the inner boundary of the accretion disc. In case of DSB and RZ
viscosity this singular point is hyperbolic and can be classified as an improper
node (see Fig.~\ref{fig:phasediagram_node}). Its characteristic directions are the
$x$-axis along which infinitely many solutions approach the singular point and the
straight line
\begin{equation}
  \label{eqn:no_torque_solutions_DSB_RZ}
  y(x) = \left(\tfrac{3}{2}\right)^j\,\frac{x+\tfrac{3}{2}}{\kappa+\tfrac{3}{2}}
  \quad\textnormal{with}\quad
  j = \begin{cases}
        1\quad\textnormal{for DSB} \\
        2\quad\textnormal{for RZ.}
      \end{cases}
\end{equation}
Along this line there exist exactly two distinct solutions approaching the
singular point from the two opposing directions. If $\kappa=-\tfrac{3}{2}$ it
becomes the vertical line $x=-\tfrac{3}{2}$.

The characteristic directions separate the reasonable solutions from those which
contradict the requirements mentioned previously. Therefore only solutions from
the upper right quadrant which pass a point between the characteristic
directions (shaded region in Fig.~\ref{fig:phasediagram_node}) will proceed to
the singular point with $y>0$ and $x>-\tfrac{3}{2}$. The analysis of the
linearized problem reveals that
\begin{equation*}
  y(x) \propto \left(x+\tfrac{3}{2}\right)^\frac{3}{2}
\end{equation*}
for all solutions approaching the singular point along the $x$-axis whereas the
solution along the other characteristic direction follows the straight line given
by Eq.~(\ref{eqn:no_torque_solutions_DSB_RZ}). All these solutions obey
$x\to-\tfrac{3}{2}$ as $y\to 0$. Since $-x$ is the local power law exponent of
$y(\xi)$ (see Eq.~\ref{eqn:selfsim_ode2}) one concludes that
$y\propto\xi^\frac{3}{2}$ for small $y$ or inversely $\xi\propto y^\frac{2}{3}$
as $\xi\to 0$. Hence we can compute the limit in the boundary condition given
by Eq.~(\ref{eqn:bc_vanishing_torque})
\begin{equation*}
  \lim\limits_{\xi\to0} \xi^5\,y^{-4}\,z(x) \propto
  \lim\limits_{x\to-\frac{3}{2}} \left(x+\tfrac{3}{2}\right)^q z(x)
\end{equation*}
where $q=-\tfrac{2}{3}$ for the distinct solution which follows the straight
line and $q=-1$ for all other solutions approaching the singular point along the
$x$-axis. We infer from the definition of $z(x)$ in Eq.~(\ref{eqn:definition_zx})
that the limit is zero (vanishing torque) only for $q>-1$ and finite for $q=-1$.
Hence there exists a unique solution which matches the no-torque condition at
the inner boundary, namely the solution with $q=-\tfrac{2}{3}$. Moreover, there
are infinitely many solutions each of them associated with a distinct function
describing the time dependence of the torque at the inner boundary.

\begin{figure}
  \centering
  \includegraphics[width=\linewidth]{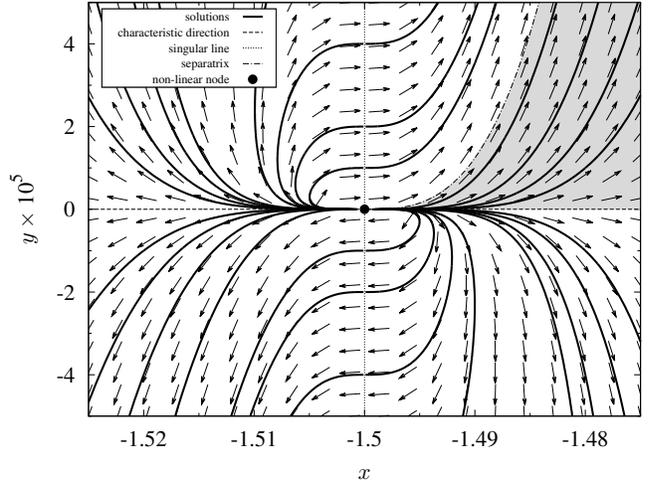}
  \caption{Phase diagram around the singular point $\left(-\tfrac{3}{2},0\right)$
     for LP viscosity prescription with $\kappa=-1$; values on the $y$-axis are
     scaled by a factor of $10^5$.}
  \label{fig:phasediagram_node_lp}  
\end{figure}
If we change over to the LP viscosity prescription we observe that the shape of
the integral curves near the singular point changes decisively (see
Fig.~\ref{fig:phasediagram_node_lp}). The singular point is no longer
hyperbolic because all linear terms of the planar system vanish identically.
Instead of that it becomes a genuinely non-linear node with a single
characteristic direction along the $x$-axis. In addition there is another
singular line at $x=-\tfrac{3}{2}$ where the derivative of $x$ with respect
to $\ln\xi$ tends to infinity.

We must draw our attention again to the upper right quadrant where
$x>-\tfrac{3}{2}$ and $y>0$. There exist two families of integral curves which 
differ in there asymptotic behavior when approaching the singular line at
$x=-\tfrac{3}{2}$. Solutions that belong to the first family reach the singular
line at a finite $y$-value and those from the other family approach the singular
point at $\left(-\tfrac{3}{2},0\right)$. The latter reside inside the shaded
area in Fig.~\ref{fig:phasediagram_node_lp}. These are the only solutions which
are permitted.

\begin{figure}
  \centering
  \includegraphics[width=\linewidth]{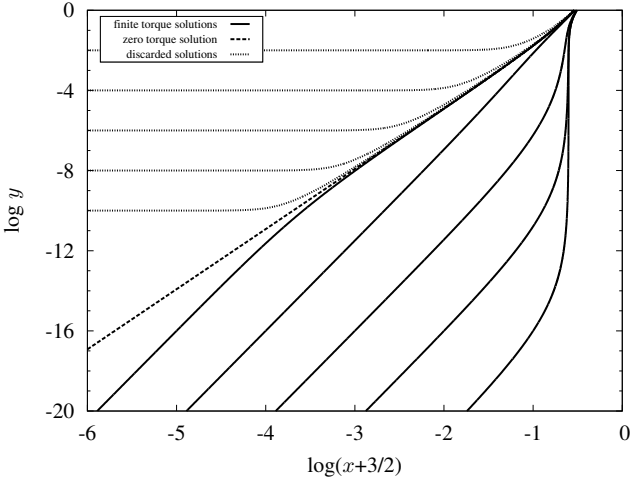}
  \caption{Log-log plot of the two families of integral curves for LP viscosity
    prescription with $\kappa=-1$; the dashed line shows the unique solution
    obtained for the no-torque boundary condition.}
  \label{fig:solutions_inv_node_lp_loglog}  
\end{figure}

In contrast to the linear analysis discussed above the separatrix between these
curves is no longer a straight line -- not even in the immediate vicinity of the
singular point. Its precise progression is unknown and cannot be deduced
analytically, because this would involve solving the non-linear problem.
Nevertheless one can obtain some important information from the non-linear
analysis\footnote{We omit the details here and refer to \cite{frommer1928} for
an elaborate discussion of the problem.}. First of all there exists again an
unique solution (dashed line in Fig.~\ref{fig:solutions_inv_node_lp_loglog}) 
which approaches the singular point along the line given by the cubic function
\begin{equation}
  \label{eqn:no_torque_solution_LP}
  y(x)=\frac{6}{\kappa+\frac{3}{2}} \left(x+\tfrac{3}{2}\right)^3.
\end{equation}
In the vicinity of the singular point all solutions which pass any point above
this curve will not proceed to the singular point. Instead they will end
somewhere on the singular line as $x\to-\tfrac{3}{2}$ with $y>0$. Secondly,
all integral curves between the unique solution and the characteristic
direction $y=0$ approach the singular point along a curve with
\begin{equation*}
  y(x) \propto \left(x+\tfrac{3}{2}\right)^\frac{9}{2}.
\end{equation*}
There are infinitely many solutions with this asymptotic behavior differing
only in their constant of proportionality.

If we take these results and insert them into the boundary condition
Eq.~(\ref{eqn:bc_vanishing_torque}) in the same way as it was done in the
linear case, one finds that the exponent $\tfrac{9}{2}$ is compatible with
the finite torque boundary condition whereas the power law in
Eq.~(\ref{eqn:no_torque_solution_LP}) would cause the torque to vanish at the
inner boundary.

In summary, it can be stated that a feasible solution $y(x)$ of the self-similar
disc problem is uniquely determined by the parameter $\kappa$ and the torque
applied to the disc at the inner boundary. The remaining problem is to specify
the constant of integration in Eq.~(\ref{eqn:selfsim_ode3}) which uniquely
determines $y$ as a function of the similarity parameter $\xi$ and hence
$\Omega(r,\tau)$. This can be achieved in principle by fixing the constant
related to the initial condition (Eq.~\ref{eqn:initial_condition_nondim}).
Unfortunately this would imply that we have to fix the asymptotic behaviour
as $\xi\to\infty$ in addition to the condition already applied at the inner
boundary as $\xi\to 0$. This would turn the initial value problem into a
boundary value problem which is slightly more difficult to treat numerically.
Instead we proceed in a different way which allows us to apply all auxiliary
conditions at the inner boundary.

\subsubsection{Solution procedure}
\label{sec:solution_procedure}

It was already mentioned above that all feasible solutions approach the
singular point at $x=-\frac{3}{2}$. These solutions converge towards
\begin{equation}
  \label{eqn:yofxi_asymptotic}
  y(\xi) = y_0\xi^\frac{3}{2}
\end{equation}
as $\xi\to 0$ where $y_0$ is some not yet determined constant which is related
to the non-dimensionalized central mass
\begin{equation*}
  \Mc(\tau) = \lim_{r\to0}M
  = \lim_{r\to0}r^3\Omega^2.
\end{equation*}
If we substitute $r$ and $\Omega$ using the group invariants (Eq.~\ref{eqn:group_invariants})
this could be written in terms of $\xi$ and $y(\xi)$ according to
\begin{equation}
  \label{eqn:central_mass}
  \Mc(\tau) = \frac{\tau^{-\left(\frac{3}{\kappa}+2\right)}}{\kappa^2}
    \lim_{\xi\to0}\frac{\xi^3}{y^2}
  = \frac{\tau^{-\left(\frac{3}{\kappa}+2\right)}}{\kappa^2y_0^2}
%  = \Mc(\tau_0) \left(\frac{\tau}{\tau_0}\right)^{-\left(\frac{3}{\kappa}+2\right)}
\end{equation}
where we used Eq.~(\ref{eqn:yofxi_asymptotic}) to compute the limit. This
remarkable result is not only an analytic formula for the exact temporal
evolution of the central mass, but allows us also to compute the unknown
constant $y_0$ once we specify the central mass $\Mc$ at some time $\tau_0$.

\begin{table}
\centering
\begin{tabular}{lcc}
\toprule
~&\makebox[3em]{DSB/RZ}&\makebox[3em]{LP}\\
\midrule
zero torque\smallskip & $1$ & $3$ \\
finite torque & $\tfrac{3}{2}$ & $\tfrac{9}{2}$\\
\bottomrule
\end{tabular}
\caption{Value of the exponent $q$ for different viscosity prescriptions and
boundary conditions.}
\label{tab:exponent_q}
\end{table}
In the previous section we showed that in the limit $x\to -\tfrac{3}{2}$ the 
function $y$ depends on $x$ according to
\begin{equation}
  \label{eqn:yofx_asymptotic}
  y(x) = y_1\left(x+\tfrac{3}{2}\right)^q
\end{equation}
where the exponent $q$ takes different values depending on viscosity prescription
and boundary condition (see Tab.~\ref{tab:exponent_q}). The constant $y_1$ has
to be determined from the inner boundary condition. In case of vanishing torque
we already derived the values for $y_1$ in Eqs.~(\ref{eqn:no_torque_solutions_DSB_RZ})
and (\ref{eqn:no_torque_solution_LP}) for all three viscosities. For the 
finite torque boundary condition one has to compute the limit of
Eq.~(\ref{eqn:viscous_torque_nondim}) as $r\to 0$ which yields
\begin{equation}
  \label{eqn:inner_torque}
  G_\star(\tau)=\lim_{r\to 0} G(r,\tau) = -\frac{\beta\zeta}{\kappa^4}
  y_0^{-\frac{10}{3}} y_1^{-\frac{2}{3}} \tau^{-\left(\frac{5}{\kappa}+4\right)}
\end{equation}
where the constant factor $\zeta$ depends on viscosity ($3$ for DSB;
$\tfrac{9}{2}$ for RZ; $12$ for LP). We can now compute the value of $y_1$, if we
specify a distinct value for the torque $G_\star$ on the inner rim at some time
$\tau_0$ provided that we have already determined $y_0$ from condition
(\ref{eqn:central_mass}).

Summing up the solution procedure there are basically six steps:
\begin{enumerate}
    \item select the viscosity parametrization
    \item specify $\beta$, $\kappa$, $\tau_0$, $\Mc(\tau_0)$, $G_\star(\tau_0)$
    \item compute $y_0$ and $y_1$
    \item choose $\xi_0\ll 1$ close to the singular point
    \item compute $y(\xi_0)$ and $x(y(\xi_0))$
    \item integrate the ODE.
\end{enumerate}
If the torque $G_\star(\tau_0)$ vanishes, the viscous coupling constant $\beta$
affects the solution only indirectly through the time variable $\tau$ (see
Eq.~\ref{eqn:disc_evolution_with_viscosity}). The value of $\xi_0$ is somewhat
arbitrary as long as $x(\xi_0)\approx-3/2$ holds. It is always possible to
scale down $\xi_0$ until its impact on the solution is smaller than numerical
errors introduced by the integration scheme.

\section{Results and Discussion}
\label{sec:results_discussion}

Before we start the discussion of the numerical results  we recall some
important properties of similarity solutions. The requirement that a solution
evolves self-similarly implies that there exists a relation between its
time variable $\tau$ and spatial variable $r$ which becomes manifest in the
definition of the similarity variable $\xi(r,\tau)$ (Eq.~\ref{eqn:group_invariants}).
Therefore the functions $x(\xi)$ and $y(\xi)$ carry information on both:
temporal evolution and spatial dependency. For a fixed instant of time $\xi$ is
simply proportional to the radial coordinate $r$ whereas for a constant radius
$\xi$ is proportional to some power of the time variable $\tau$. If one uses
logarithmic scaling this relation becomes a simple linear map between the
logarithms: $\ln\xi = \ln r+\kappa^{-1}\ln\tau$. Hence we may convert the
dependence on $\ln\xi$ into either spatial dependence on $\ln r$ applying a
constant shift or temporal evolution by scaling with $\kappa$ plus adding a
constant shift. Recall that $\kappa<0$. Therefore increasing $\xi$ corresponds
to decreasing $\tau$ and vice versa.

\subsection{Similarity solutions}
\label{sec:similarity_solutions}

\begin{figure}
  \centering
  \includegraphics[width=\linewidth]{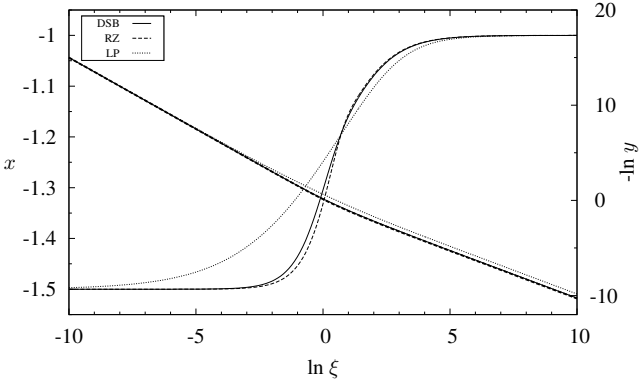}
  \caption{Zero torque solutions for different viscosity prescriptions
    obtained with $\kappa=-1,\tau_0=1,\Mc(\tau_0)=1,\ln\xi_0=-10$. The figure shows the
    results for both functions $x(\xi)$ (bottom left to top right, left axis)
    and $y(\xi)$ (top left to bottom right, right axis).}
  \label{fig:solutions_xofxi_notorque}
\end{figure}

\begin{figure}
  \centering
  \includegraphics[width=\linewidth]{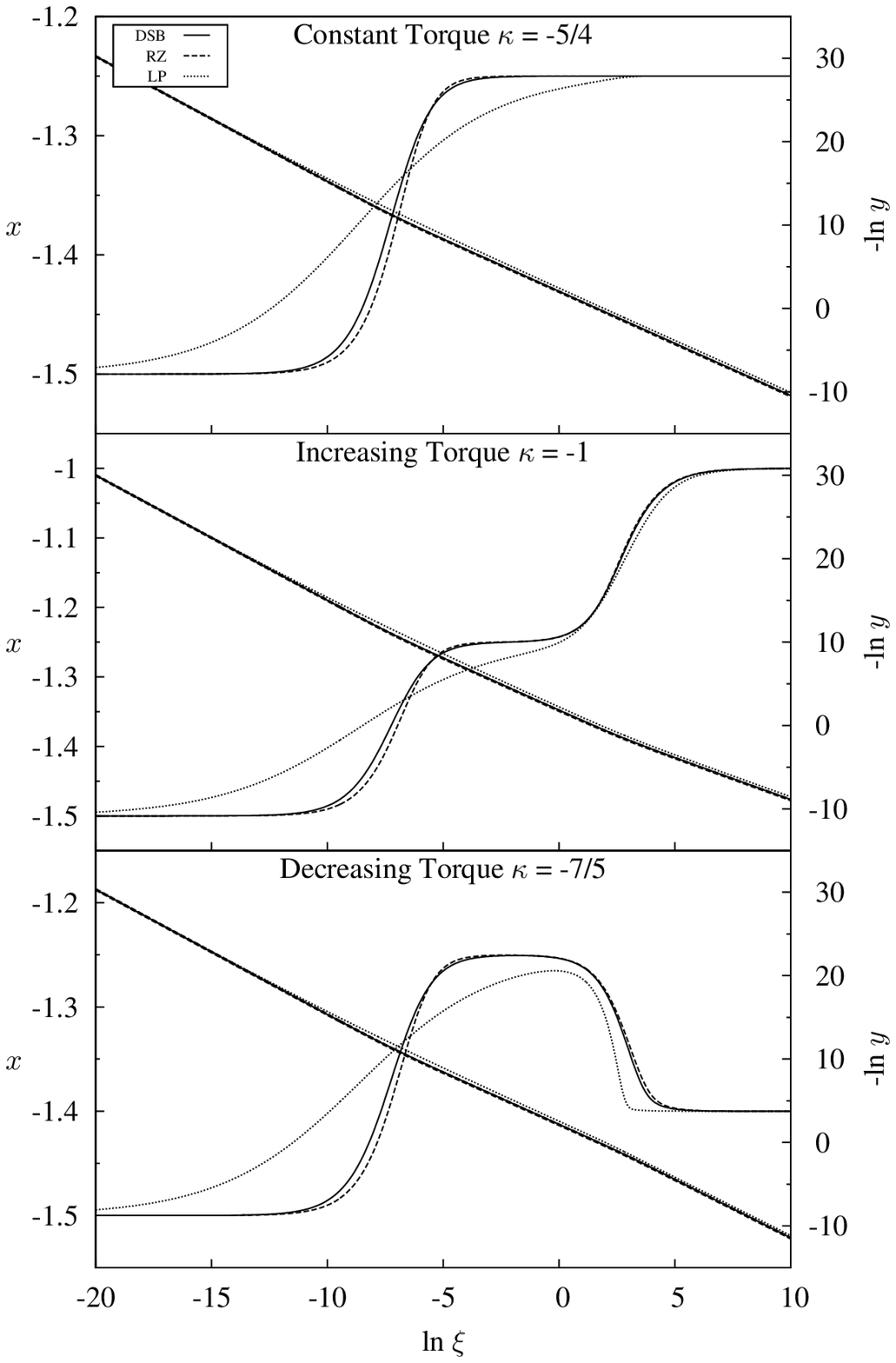}
  \caption{Finite torque solutions for different viscosity prescriptions obtained
    with $\tau_0=1,\Mc(\tau_0)=1, \beta=10^{-3}, G_\star(\tau_0)=-1,\ln\xi_0=-25$.
    The figure shows the results for both functions $x(\xi)$ (bottom left to top
    right, left axis) and $y(\xi)$ (top left to bottom right, right axis).}
  \label{fig:solutions_xofxi_withtorque}
\end{figure}

In contrast to the rather complex structures shown in the phase diagram
in Fig.~\ref{fig:phasediagram_big} the solutions look quite simple. The shape of 
$x(\xi)$ is basically a step function with a smooth transition between two
or three levels depending on the inner boundary condition. If the torque
applied to the disc at the inner boundary is constant (Fig.~\ref{fig:solutions_xofxi_withtorque}
upper panel) or zero (Fig.~\ref{fig:solutions_xofxi_notorque}) the two levels are
$x_{-\infty}=-3/2$ and $x_{+\infty}=\kappa$. In the two cases where the torque
at the inner boundary changes with time there is an additional intermediate level
at $x_0=-5/4$ (Fig.~\ref{fig:solutions_xofxi_withtorque} lower panels). Since
$-x$ is the logarithmic derivative of $y$ with respect to $\xi$
(Eq.~\ref{eqn:selfsim_ode2}) $y(\xi)$ is essentially a broken power law.

Apart from the influence of the inner boundary condition the solutions show a more
or less prominent dependence on the viscosity law. The results for DSB and RZ
viscosity are very similar as one would expect, because the viscosity prescriptions
differ by a factor of $|x|$ which is always of order one and becomes at most $3/2$
in the Keplerian limit where both solutions approach $x_{-\infty}=-3/2$. In both
cases the transition between the different constant levels of $x$ is quite sharp.
In contrast to that the results for $x(\xi)$ with LP viscosity exhibit always a 
smoother transition. In case of decreasing and increasing torque
(Fig.~\ref{fig:solutions_xofxi_withtorque} lower panels) the intermediate
level at $x_0=-5/4$ is hardly visible.

Once the similarity problem has been solved for the non-dimensional functions
$x(\xi)$ and $y(\xi)$ one can extract from these the complete information about
the self-similar evolution using the definition of the group invariants in
Eqs.~(\ref{eqn:group_invariants}). Thus with Eq.~(\ref{eqn:group_invariant_solution})
one obtains $\Omega$ as a function of $r$ and $\tau$ and with
Eqs.~(\ref{eqn:monopole_approximation}) and (\ref{eqn:sigma_omega_relation}) one
computes the non-dimensionalized\footnote{see remark in the first paragraph of
Sec.~\ref{sec:self-similar_solutions}} expressions for enclosed mass and surface
density
\begin{align}
  \label{eqn:enclosed_mass_nondim}
  M &= \kappa^{-2}\,\tau^{-\left(\frac{3}{\kappa}+2\right)}\,\xi^3\,y^{-2} \\
  \label{eqn:surface_density_nondim}
  \sden &= \tinv{2\pi}\kappa^{-2}\,\tau^{-\left(\inv{\kappa}+2\right)}\,
                    (2x+3)\,\xi\,y^{-2}.
\end{align}
The local accretion rate can be derived from (\ref{eqn:enclosed_mass_nondim})
by differentiation
\begin{equation}
  \label{eqn:accretion_rate_nondim}
  \dot{M}=\partial_t M= \beta\partial_\tau M=2\beta\kappa^{-3}\,\tau^{-\left(\frac{3}{\kappa}+3\right)}\,
                   (x-\kappa)\,\xi^3\,y^{-2}
\end{equation}
and with help of Eqs.~(\ref{eqn:enclosed_mass_gradient}) and (\ref{eqn:enclosed_mass_transport})
one obtains the radial velocity
\begin{equation}
  \label{eqn:radial_velocity_nondim}
  v_r=-\frac{\dot{M}}{2\pi r\sden} = -2\beta\kappa^{-1}\tau^{-\left(\inv{\kappa}+1\right)}\,
            \xi\,\frac{x-\kappa}{2x+3},
\end{equation}
if $\kappa\ne -3/2$. The special case where $\kappa=-3/2$ is discussed in more
detail below.

The viscous torque is given by Eq.~(\ref{eqn:viscous_torque_nondim}). In
addition we can derive an expression for the vertically integrated dissipation
rate \cite*[see][Chap.\ 3]{kato2008} using Eq.~(\ref{eqn:definition_combined_viscosity})
to substitute the viscosity
\begin{equation}
  \label{eqn:dissipation_rate}
  Q_\mathrm{vis} = \nu\sden\left(r\partial_r\Omega\right)^2=\beta r^2 \Omega^3
       \sden\,x^2 f(x).
\end{equation}
We can again utilize the group invariants (\ref{eqn:group_invariants}), insert
$\sden$ from Eq.~(\ref{eqn:surface_density_nondim}) and substitute $f(x)$ with
help of Eq.~(\ref{eqn:definition_zx}):
\begin{equation}
  \label{eqn:dissipation_rate_nondim}
  Q_\mathrm{vis} = \tfrac{\beta}{2\pi} (-\kappa)^{-5}\tau^{-\left(\frac{3}{\kappa}+5\right)}\,
        x z(x)\,\xi^{3}\,y^{-5}.           
\end{equation}
Observe that all expressions derived above depend on $\xi(r,\tau)$, explicitly
as well as implicitly through $x(\xi)$ and $y(\xi)$. Thus in addition to the
explicit dependence on $\tau$ there is an implicit dependence through the
similarity variable $\xi$.

\subsection{Self-similar time evolution}
\label{sec:time_evolution}

\begin{figure}
  \centering
  \includegraphics[width=\linewidth]{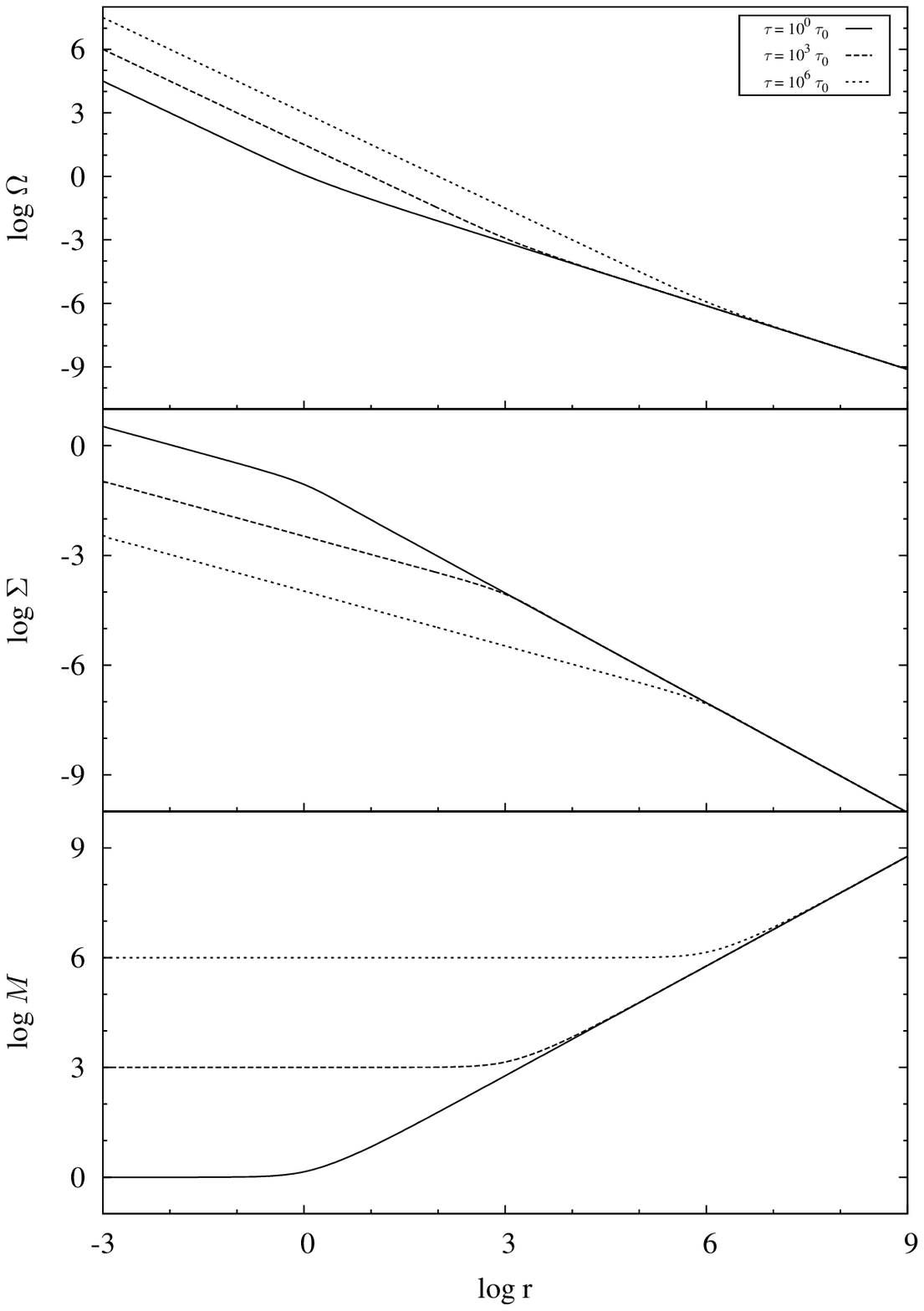}
  \caption{Self-similar solutions of angular velocity, surface density and
    enclosed mass at three different times. All quantities
    are given in non-dimensional units. The figures show results for the zero
    torque boundary condition obtained with DSB viscosity prescription and
    parameters $\kappa=-1, \beta=10^{-3}, \tau_0=1, \Mc(\tau_0)=1,\ln\xi_0=-10$.}
  \label{fig:time_evolution_dsb}
\end{figure}
\begin{figure}
  \centering
  \includegraphics[width=\linewidth]{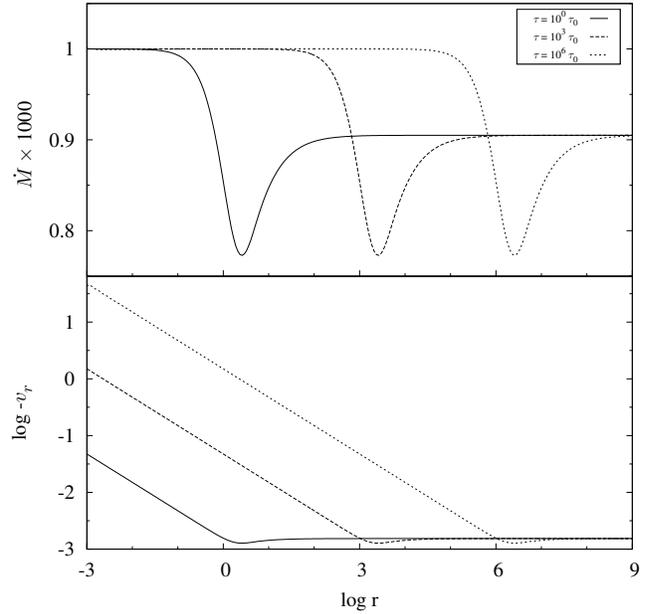}
  \caption{Self-similar solutions of accretion rate $\dot{M}$ and radial
    velocity at three different times. All quantities are given in non-dimensional
    units. The figures show results for the zero torque boundary condition obtained
    with DSB viscosity prescription and parameters $\kappa=-1, \beta=10^{-3},
    \tau_0=1, \Mc(\tau_0)=1,\ln\xi_0=-10$.}
  \label{fig:time_evolution_dsb1}
\end{figure}
In figures \ref{fig:time_evolution_dsb} and \ref{fig:time_evolution_dsb1} we
show the numerical results of the $\kappa=-1$ solution obtained with zero torque
boundary condition and DSB viscosity prescription. At any fixed time the
radial dependence of any function is basically a broken power law. The kink
thereby separates the region in which the central object dominates the gravitational
potential from that in which self-gravity of the disc becomes important. One can
obtain the asymptotic exponents of these power laws analytically from the expressions
given above using the asymptotic expansion of the functions $x(\xi)$ and $y(\xi)$.
The exponents are listed in Tab.~\ref{tab:asymptotic_exponents_radial}.

In addition to the spatial dependence we show the numerical results for three
different values of the time variable $\tau$ in the diagrams. The temporal
evolution is given by a simple shift in the log-log diagrams. Thereby the radial
shift is determined by the definition of the similarity variable $\xi$ 
(Eq.~\ref{eqn:group_invariants}) which relates time dependence to spatial
dependence. Hence the kink where the power law exponent changes is shifted by a
factor of $-\inv{\kappa}\log(\frac{\tau_2}{\tau_1})$ on the logarithmic scale if
time progresses from $\tau_1$ to $\tau_2$. In the limit $r\to 0$ we can compute
analytical expressions for the time dependence which
can again be expressed in terms of power laws. The exponents are listed in 
Tab.~\ref{tab:asymptotic_exponents_temporal}. In the limit of large radii it is
not surprising that there is obviously no temporal evolution because the time
scales become larger and the solutions do not evolve at all.

We would like to emphasize that except for surface density and radial velocity the power law
exponents of the asymptotic expansions do not depend on the viscosity prescription.
Furthermore angular velocity, enclosed mass and accretion rate show always the
same asymptotic progression no matter what kind of viscosity prescription or inner
boundary condition has been selected. In the Keplerian limit enclosed mass and
accretion rate become almost constant with respect to radial distance. Thus we
retain an important feature of the standard steady disc models \cite[see, \eg][]{pringle1981}
in this limit. However, since we account for mass accretion on to the central
object, it inevitably becomes more massive as time progresses. Therefore the
central mass as well as the accretion rate generally depend on time. There are
two exceptions namely the solutions obtained for $\kappa=-3/2$ and $\kappa=-1$
which we discuss below.

In the standard theory of steady state accretion discs the viscous dissipation
rate  $Q_\mathrm{vis}$ is proportional to the central mass multiplied by the
accretion rate and divided by $r^3$ \cite[see, \eg][]{pringle1981}\footnote{There
is usually another multiplicative factor due to boundary conditions depending on
$r$ as well which we ignore here.}. The viscous dissipation rate obtained for
our solutions with zero torque inner boundary condition shows exactly the same
asymptotic progress for small radii. This becomes clearer if we express 
$Q_\mathrm{vis}$ in terms of accretion rate and enclosed mass. With help of
Eqs.~(\ref{eqn:definition_combined_viscosity}), (\ref{eqn:sigma_omega_relation})
and (\ref{eqn:accretion_rate_nondim}) one can express the product of viscosity
and surface density in terms of the accretion rate and the similarity solution
$x(\xi)$ and $y(\xi)$:
\begin{equation*}
  \nu\sden = -\frac{\dot{M}}{4\pi}\,\frac{(2x+3)f(x)}{(x-\kappa)y}.
\end{equation*}
If one substitutes $\nu\sden$ in Eq. (\ref{eqn:dissipation_rate}) and uses the
balance law (\ref{eqn:monopole_approximation}) to eliminate $\Omega^2$ from the
equation, we can express the dissipation rate in terms of accretion rate,
enclosed mass and the similarity solution $x(\xi)$ and $y(\xi)$ (recall that
we use non-dimensional functions and therefore set the gravitational constant
$G=1$):
\begin{equation}
  \label{eqn:dissipation_rate_mdot_mass}
  Q_\mathrm{vis} = -\frac{\dot{M}M}{4\pi r^3}\,\frac{xz(x)}{(x-\kappa)y}.
\end{equation}
Thereby we utilized the viscosity dependent function $z(x)$ defined in
Eq.~(\ref{eqn:definition_zx}). One can easily proof in case of the zero torque
boundary condition that the second fraction in Eq.~(\ref{eqn:dissipation_rate_mdot_mass})
becomes
\begin{equation*}
   \lim_{x\to-\frac{3}{2}} \frac{xz(x)}{(x-\kappa)y} = -3
\end{equation*}
in the Keplerian limit for all three viscosity prescriptions. Thus we can
recover an important relation known from steady state accretion disc theory.
However, in our case $\dot{M}$ and $M$ are time dependent functions in the
limit $r\to 0$ (see Tab.~\ref{tab:asymptotic_exponents_temporal}) and therefore
$Q_\mathrm{vis}$ is also time dependent. If we examine solutions obtained with
the finite torque boundary condition we yield slightly steeper radial slopes
(see Tab.~\ref{tab:asymptotic_exponents_radial}) for the dissipation rate if
$r\to 0$.

In addition we observe a considerable deviation from the $r^{-3}$ law depending
on the value of $\kappa$ at large radii where the disc becomes fully
self-gravitating. This is not a surprising result, because the material in
the disc contributes to the gravitational potential energy. Hence there exists
an additional energy source which alters the radial dependence of the viscous
dissipation rate.
\begin{table}
\centering
\begin{threeparttable}
\begin{tabular}{lccccc}
\toprule
& \multicolumn{4}{c}{$r\to 0$} & $r\to\infty$ \\
b.\ c.\ & \multicolumn{2}{c}{zero torque} & \multicolumn{2}{c}{finite torque} & \\
vis.\ & \scriptsize{DSB/RZ} & \scriptsize{LP} & \scriptsize{DSB/RZ} & \scriptsize{LP} & \\
\midrule
$\Omega$  & \multicolumn{4}{c}{$-\frac{3}{2}$} & $\kappa$ \\[1.5ex]
$\sden$ & $-\frac{1}{2}$ & $-\frac{3}{2}$ & $-1$ & $-\frac{5}{3}$ & $2\kappa+1$\tnote{$\dagger$} \\[1.5ex]
$M$       & \multicolumn{4}{c}{$0$} & $2\kappa+3$ \\[1.5ex]
$\dot{M}$ & \multicolumn{4}{c}{$0$} & $3\kappa+3$\tnote{$\dagger$,$\star$} \\[1.5ex]
% $J$ & $0$ & $-1$ & $-\frac{1}{2}$ & $-\frac{7}{6}$ & $3\kappa+3$\tnote{$\dagger$} \\[1.5ex]
$v_r$\tnote{$\ddagger$} & $-\frac{1}{2}$ & $\frac{1}{2}$ & $0$ & $\frac{2}{3}$ & $\kappa+1$\tnote{$\star$} \\[1.5ex]
$G$       & \multicolumn{2}{c}{$\inv{2}$} & \multicolumn{2}{c}{$0$} & $4\kappa+5$\tnote{$\dagger$}\\[1.5ex]
$Q_\mathrm{vis}$ & \multicolumn{2}{c}{$-3$} & \multicolumn{2}{c}{$-\frac{7}{2}$} & $5\kappa+3$\tnote{$\dagger$}\\
\bottomrule
\end{tabular}
\begin{tablenotes}
  \item[\quad $\dagger$] \scriptsize{if$\,\kappa=-3/2\quad$ DSB/RZ: exponential decay, LP: no solution}
  \item[\quad $\ddagger$] \scriptsize{if$\,\kappa=-3/2\quad$ zero torque: $v_r=0$, finite torque: 
     see Eq.~(\ref{eqn:radial_velocity_constant_mass_solution})}
  \item[\quad $\star$] \scriptsize{if$\,\kappa=-5/4\quad$ exponential decay}
\end{tablenotes}
\end{threeparttable}
\caption{Asymptotic behaviour of angular velocity, surface density, enclosed mass,
accretion rate, radial velocity, viscous torque and dissipation rate for different inner boundary
conditions and viscosity prescriptions. The table lists the power law exponents
of the radial dependence for small and large radii.}
\label{tab:asymptotic_exponents_radial}
\end{table}
\begin{table}
\centering
\begin{threeparttable}
\begin{tabular}{lcccc}
\toprule
b.\ c.\ & \multicolumn{2}{c}{zero torque} & \multicolumn{2}{c}{finite torque} \\
vis.\ & \scriptsize{DSB/RZ} & \scriptsize{LP} & \scriptsize{DSB/RZ} & \scriptsize{LP} \\
\midrule
$\Omega$  & \multicolumn{4}{c}{$-\left(\frac{3}{2\kappa}+1\right)$} \\[1.5ex]
$\sden$ & $\left(\frac{-3}{2\kappa}-2\right)$ & $\left(\frac{-5}{2\kappa}-2\right)$ & $\left(\frac{-2}{\kappa}-2\right)$ & $\left(\frac{-8}{3\kappa}-2\right)$ \\[1.5ex]
$M$       & \multicolumn{4}{c}{$-\left(\frac{3}{\kappa}+2\right)$} \\[1.5ex]
$\dot{M}$ & \multicolumn{4}{c}{$-\left(\frac{3}{\kappa}+3\right)$\tnote{$\dagger$}} \\[1.5ex]
$v_r$\tnote{$\ddagger$} & $\left(\frac{-3}{2\kappa}-1\right)$ & $\left(\frac{-1}{2\kappa}-1\right)$ & $\left(\frac{-1}{\kappa}-1\right)$ & $\left(\frac{-1}{3\kappa}-1\right)$ \\[1.5ex]
$G$       & \multicolumn{2}{c}{$-\left(\frac{9}{2\kappa}+4\right)$} & \multicolumn{2}{c}{$-\left(\frac{5}{\kappa}+4\right)$} \\[1.5ex]
$Q_\mathrm{vis}$ & \multicolumn{2}{c}{$-\left(\frac{6}{\kappa}+5\right)$} & \multicolumn{2}{c}{$-\left(\frac{13}{2\kappa}+5\right)$} \\
\bottomrule
\end{tabular}
\begin{tablenotes}
  \item[\quad $\dagger$] \scriptsize{if$\,\kappa=-3/2\quad M_\star$ is constant
       and therefore $\dot{M}$ vanishes for $r\to 0$}
  \item[\quad $\ddagger$] \scriptsize{if$\,\kappa=-3/2\quad$ zero torque: $v_r=0$, finite torque: 
     see Eq.~(\ref{eqn:radial_velocity_constant_mass_solution})}
\end{tablenotes}
\end{threeparttable}
\caption{Asymptotic behaviour of angular velocity, surface density, enclosed mass,
accretion rate, radial velocity, viscous torque and dissipation rate for different inner boundary
conditions and viscosity prescriptions. The table lists the power law exponents
of time evolution in the limit $r\to 0$.}
\label{tab:asymptotic_exponents_temporal}
\end{table}

One easily verifies that the numerical results shown in Fig.~\ref{fig:time_evolution_dsb}
and \ref{fig:time_evolution_dsb1} for the $\kappa=-1$ solution satisfy exactly
the predicted asymptotic behaviour listed in Tab.~\ref{tab:asymptotic_exponents_radial}
and \ref{tab:asymptotic_exponents_temporal}.
\Eg with those power law exponents we conclude that $M\propto r^0\tau^1$ for
$r\to 0$. Thus $M$ becomes independent of $r$ for small radii resembling the
fact that $M$ approaches the value of the central mass $\Mc$ which grows linear
with respect to the time variable $\tau$. Therefore it increases by a factor of
$10^3$ if $\tau$ increases by the same factor (see Fig.~\ref{fig:time_evolution_dsb}
lower panel).

An exceptional feature of the $\kappa=-1$ solution is the almost constant
accretion rate $\dot{M}$ (see Fig.~\ref{fig:time_evolution_dsb1} upper panel).
This is always fulfilled in the limit $r\to 0$ independent of the actual value
of $\kappa$ whereas only for $\kappa=-1$ one obtains also a constant
accretion rate for $r\to\infty$ (see Tab.~\ref{tab:asymptotic_exponents_radial}).
Although the accretion rate is almost constant with respect to both time and
radial distance the solution is not stationary. All other quantities show a
clear time dependence.

\begin{figure}
  \centering
  \includegraphics[width=\linewidth]{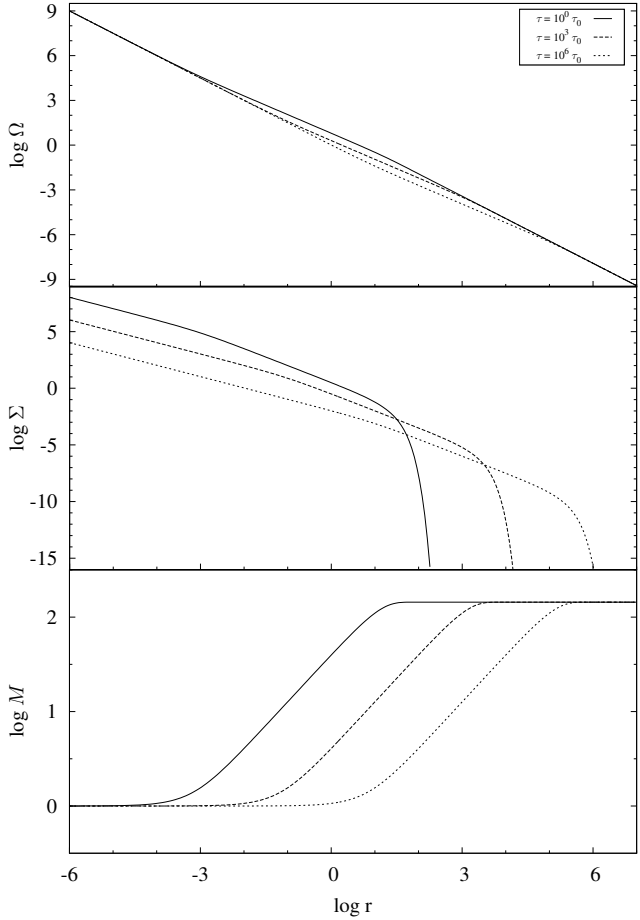}
  \caption{Self-similar time evolution of angular velocity, surface density and
    enclosed mass for the constant disc mass solution. All quantities are given in
    non-dimensional units. The figures show results for decreasing torque boundary
    condition obtained with DSB viscosity prescription and parameters
    $\kappa=-3/2, \beta=10^{-3}, \tau_0=1, \Mc(\tau_0)=1, G_\star(\tau_0)=-1,
    \ln\xi_0=-20$.}
  \label{fig:time_evolution_dsb2}
\end{figure}
Another interesting example is that obtained for $\kappa=-3/2$ which has the
unique property that the mass of the disc is finite and time-independent 
(see Sec.~\ref{sec:conservation_conditions}). As in the other cases there are two
different classes of solutions\footnote{Both solutions are prohibited by the LP
viscosity prescription, because there is a singular line at $x=-3/2$ in the phase
diagram (see Fig.~\ref{fig:phasediagram_node_lp})} depending on the torque applied
to the disc at the inner boundary. If the torque vanishes one obtains a pure Keplerian
rotation law with $x=-3/2$ everywhere at any time. Hence the enclosed mass $M$ vanishes
except for the central mass $\Mc$. However, if the torque at the inner boundary
is finite, there exists another solution with Keplerian rotation at small radii
as well as larger radii and an intermediate flatter rotation law where $x$
approaches $-5/4$ (see Fig.~\ref{fig:time_evolution_dsb2} upper panel).
In both cases there is no accretion on to the central object. In the former case
there is no mass flow at all whereas in the latter case the torque applied to
the discs inner boundary causes radial outflow. The radial velocity
given by Eq.~(\ref{eqn:radial_velocity_nondim}) with $\kappa=-3/2$ is positive
for all radii $r>0$:
\begin{equation}
  \label{eqn:radial_velocity_constant_mass_solution}
  v_r = \frac{2}{3}\beta\,\frac{r}{\tau} = \frac{2}{3}\,\frac{r}{t}.
\end{equation}
The surface density depicted in
Fig.~\ref{fig:time_evolution_dsb2} for the solution with constant disc mass and
decreasing torque shows an exponential decay\footnote{This might be a problem,
because it contradicts one of our basic assumptions (see Sec.~\ref{sec:slow_accretion_limit}).}
at large radii in contrast to all other self-similar solutions which decline
according to a power law (see Tab.~\ref{tab:asymptotic_exponents_radial}). Hence
the disc has a sharp outer rim which moves further outwards as time increases.
The disc mass is redistributed yielding a dispersion of the hole disc. The same
behaviour was also found by \cite{lin1987} if they assume in their model, which
does not account for self-gravity that the disc mass is time-independent.

\subsection{The impact of the power law exponent $\kappa$}
\label{sec:impact_of_kappa}

In Sec.~\ref{sec:auxiliary_conditions} we showed that $\kappa$ is the power
law exponent of the rotation curve at large radii. Since $\Omega$ is related to
the enclosed mass (Eq.~\ref{eqn:monopole_approximation}) whose gradient
determines the surface density distribution (Eq.~\ref{eqn:enclosed_mass_gradient}),
$\kappa$ also controls the radial behaviour of these at large radii (see
Tab.~\ref{tab:asymptotic_exponents_radial}). A steeper rotation law causes a
steeper radial decline of the surface density which in turn leads to a flatter
radial increase of the enclosed mass. Thus $\kappa$ is just another measure
of the matter distribution within the disc and hence how self-gravitating the
disc becomes at large radii.

On the other hand we found that $\kappa$ has an impact on the temporal evolution
of the disc at small radii as well. From the values of Tab.~\ref{tab:asymptotic_exponents_temporal}
we can draw some remarkable conclusions. First, the central mass always increases
with time except for the unique solution with $\kappa=-3/2$ where the central
mass remains constant. Second, the accretion rate at small radii decreases with time
if $\kappa<-1$, remains constant during the whole evolution only if $\kappa=-1$
and increases with time if $\kappa>-1$. The slope of the temporal decline
becomes flatter as $\kappa$ approaches the value $-1$. Hence, discs with higher
values of $\kappa$ evolve faster than those for which $\kappa$ is close to the
Keplerian value of $-3/2$ or thinking in terms of mass accretion rates: Objects
embedded in self-gravitating discs with flatter rotation laws grow faster than
those embedded in nearly Keplerian discs.

\begin{figure}
  \centering
  \includegraphics[width=\linewidth]{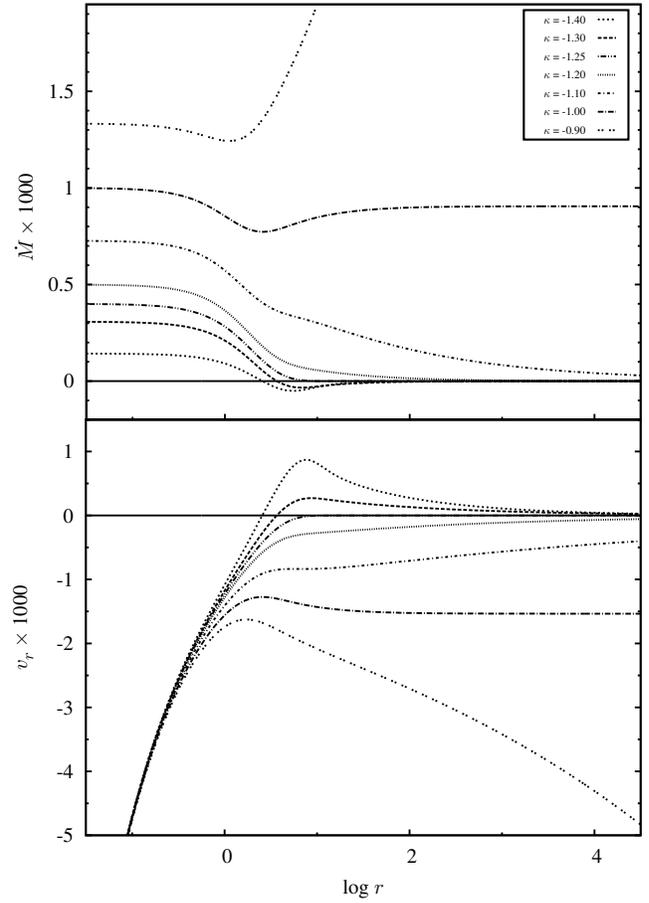}
  \caption{Accretion rate $\dot{M}$ and radial velocity $v_r$ as a function of
    radial distance $r$ at time $\tau=\tau_0$ for different
    values of the parameter $\kappa$. The values on the $y$-axis are scaled by
    a factor of $\beta^{-1}=1000$.
    The figures show results for zero torque boundary condition obtained
    with DSB viscosity prescription and parameters $\beta=10^{-3}, \tau_0=1,
    \Mc(\tau_0)=1, \ln\xi_0=-10$.}
  \label{fig:dotM_vr_dsb_notorque}
\end{figure}

If we take a look at the radial dependence of the accretion rate (Fig.~\ref{fig:dotM_vr_dsb_notorque}),
we can identify another difference between solutions obtained with different
values of the parameter $\kappa$. We already mentioned before that only for
$\kappa=-1$ the accretion rate becomes independent of radial distance in the
limit $r\to\infty$. If $\kappa$ is above that value, the accretion rate increases
as $r$ increases, it tends to zero if $\kappa<-1$ and it remains positive for all
radii if $\kappa\geq-5/4$. However, if $\kappa<-5/4$, the accretion rate falls
below zero in the transitional region between the inner Keplerian disc
and the outer self-gravitating part of the disc. The same behaviour can be seen
more clearly if we take a look at the radial velocity depicted in the lower panel
of Fig.~\ref{fig:dotM_vr_dsb_notorque}. For small radii the radial velocity is
always negative leading to the positive accretion rate in the region where the
disc is nearly Keplerian, but as $r$ increases the radial velocity becomes
positive for the solutions obtained with $\kappa=-1.4$ and $\kappa=-1.3$. These
accretion discs may loose a considerable amount of mass by redistributing it to
the outer disc rather than accreting it on to the central object. This supports
the proposition stated in the previous paragraph.

% Another important quantity which varies significantly depending on the value of
% $\kappa$ is the growth rate of enclosed angular momentum. If we define the
% enclosed angular momentum within radial distance to the centre analogous to the
% enclosed mass by
% \begin{equation}
%   \label{eqn:enclosed_angular_momenum}
%   J = 2\pi\int_0^r \sden \sam r\di{r},
% \end{equation}
% then radial integration of the angular momentum equation (\ref{eqn:angular_momentum})
% yields 
% \begin{equation}
%   \label{eqn:angular_momentum_flux}
%   \dot{J} = \partial_t J = G(t,r)-G_\star(t) + \sam\dot{M}.
% \end{equation}
% Thus there are three different contributions which cause a change in angular
% momentum: The viscous torques acting at radius $r$ and at the inner boundary plus
% the angular momentum flux associated with the mass flow at radius $r$. The torques
% we consider in our models are negative, because the angular velocity always
% decreases as $r$ increases. Hence $G(r,t)$ always reduces $\dot{J}$ whereas
% $-G_\star(r)$ increases $\dot{J}$. On the other hand the contribution of the
% third therm $\sam\dot{M}$ depends on the sign of the local accretion rate $\dot{M}$
% which can be either positive (mass inflow) or negative (mass outflow).

\subsection{Verification of the model assumptions}
\label{sec:verify_assumptions}

The derivation of the disc evolution equation in Sec.~\ref{sec:disc_equation}
relies on two essential assumptions, namely the thin disc approximation which
is related to the supersonic rotation requirement (see Eq.~\ref{sec:supersonic_rotation})
and the slow accretion limit. The verification of the former would require
knowledge about the speed of sound and therefore the temperature in the midplane
of the disc (see Eq.~\ref{eqn:disc_height}). Hence in order to check for
supersonic rotation, one generally has to solve the vertical structure problem
which is beyond the scope of this work.\footnote{We explain in Section~\ref{sec:application}
how to compute a rough estimate of the central temperature for a typical AGN
accretion disc.}

Nevertheless we can compute the ratio $\vs/\vphi$ and figure out if the radial
velocity is always much smaller than the azimuthal velocity. This is at least a
necessary condition for the slow accretion limit. From Eqs.~(\ref{eqn:group_invariant_solution})
and (\ref{eqn:radial_velocity_nondim}) one easily computes
\begin{equation}
  \label{eqn:velocity_ratio}
  \frac{\vs}{\vphi} = \beta\,\frac{(x-\kappa)\,y}{x+\frac{3}{2}}.
\end{equation}
Hence $\vs/\vphi$ scales with the non-dimensional viscous coupling constant $\beta$
and depends on the non-dimensional functions $x(\xi)$ and $y(\xi)$. We can use the
numerically obtained similarity solutions to compute the inverse $\xi(x)$ and insert
the result in $y(\xi)$ to express the right hand side of Eq.~(\ref{eqn:velocity_ratio})
in terms of $x$. This allows us to express the velocity ratio as a function of
$x$ alone.

\begin{figure}
  \centering
  \includegraphics[width=\linewidth]{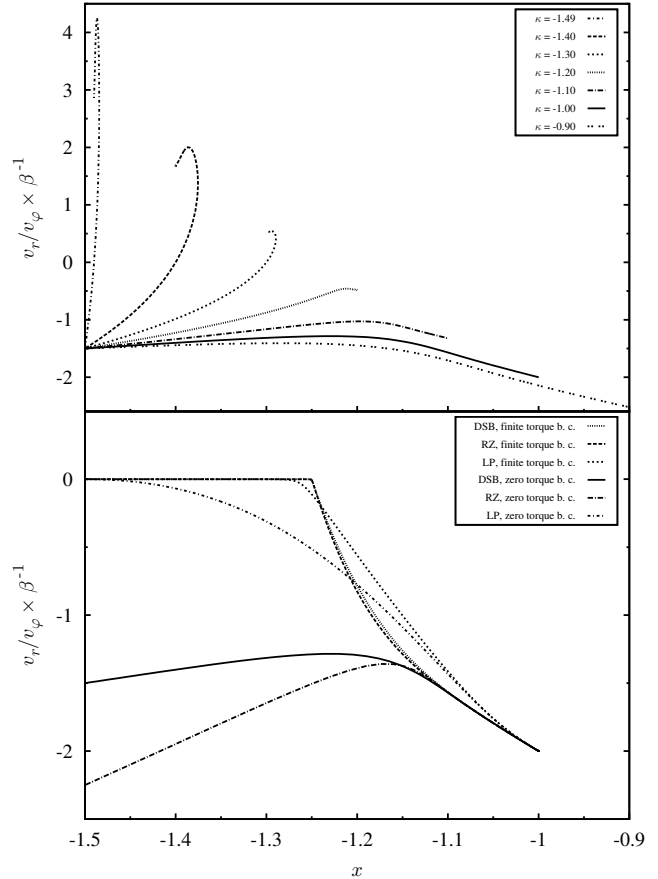}
  \caption{Ratio of radial and azimuthal velocity $\vs/\vphi$ as a function of
    local power law exponent $x$. The top panel shows results obtained with DSB
    viscosity and zero torque boundary condition for different values of the
    parameter $\kappa$. The lower panel shows solutions for different viscosity
    prescriptions and boundary conditions for $\kappa=-1$. The values on the
    $y$-axis are scaled by a factor of $\beta^{-1}=1000$. The other simulation
    parameters are $\tau_0=1, \Mc(\tau_0)=1, \ln\xi_0=-10$ for zero torque
    solutions and $\ln\xi_0=-25$ for finite torque solutions with
    $G_\star(\tau_0)=-1$.}
  \label{fig:vrDvphi}
\end{figure}

The results for various numerical solutions obtained with different values of the
parameter $\kappa$ and with different viscosity prescriptions and boundary
conditions are depicted in Fig.~\ref{fig:vrDvphi}. One can clearly see that in all
cases $\vs/\vphi\,\beta^{-1}$ is of the order of one and therefore $\vs/\vphi$
must be of the order of $\beta$ which is usually set to values smaller than
$10^{-2}$ (see Sec.~\ref{sec:viscosity_prescription}). Therefore we conclude that
as long as $\beta\ll1$, the radial velocity $\vs$ is indeed much smaller than
the azimuthal velocity $\vphi$ and furthermore if $\cs$ is only slightly larger
than $\vs$ the azimuthal velocity must be highly supersonic.

\subsection{Application to AGN discs}
\label{sec:application}

In this section we exemplify how to introduce physical dimensions and apply them
to the non-dimensional solutions shown in the previous sections. As was already
mentioned in the introductory paragraph of Sec.~\ref{sec:self-similar_solutions}
the non-dimensionalization of the disc equations can be achieved by specifying 
two independent basic scales, \eg, length and mass. Then a third independent 
scale, \eg, time, is given by Eq.~(\ref{eqn:time_scale}). One can specify any
two of the three scales for length, mass and time and compute the remaining scale.
Once the basic scales have been determined one can derive the scales for
angular velocity and linear velocity, surface density and viscous dissipation
rate:
\begin{equation*}
  \widetilde{\Omega}=\tilde{\tau}^{-1},\quad
  \tilde{v}=\tilde{r}\tilde{\tau}^{-1},\quad
  \widetilde{\sden}=\widetilde{M}\tilde{r}^{-2},\quad
  \widetilde{Q}=\widetilde{M}\tilde{\tau}^{-3}
\end{equation*}
and since $\tau=\beta\,t$ one can compute the scale of the real time variable $t$
and the mass accretion rate
\begin{equation*}
  \tilde{t}=\beta^{-1}\tilde{\tau},\quad
  \widetilde{\dot{M}}=\widetilde{M}\tilde{t}^{-1}.
\end{equation*}
In Tab.~\ref{tab:scaling_AGN} we list the scaling constants of a self-gravitating
accretion disc model applicable to AGN. These scales can be used as units for
the non-dimensional model shown in Fig.~\ref{fig:time_evolution_dsb},
\eg the numbers on the x-axis denote the common logarithm of radial distance to
the central black hole in astronomical units (AU). \alert{To convert the radial
scale from AU to parsec (pc) one has to subtract $5.3$. Hence the outer boundary
at $10^8$\,AU corresponds to $10^{2.7}$\,pc $\approx 500$\,pc or
$1.5\cdot10^{21}$\,cm. If we set $\beta=10^{-3}$ (for a discussion of reasonable
values for $\beta$ see Sec.~\ref{sec:viscosity_prescription} and the references
given there) we can determine the constant accretion rate of the central black
hole from the upper panel of Fig.~\ref{fig:time_evolution_dsb1}:}
\begin{equation*}
  \dot{M} = \beta\,\widetilde{\dot{M}}=10^{-3}\cdot 200\,\Msun\,\yr^{-1}s
          = 0.2\,\Msun\,\yr^{-1}.
\end{equation*}
This is a quite reasonable value for the late phases of the accretion process.
However, if the mass of the central black hole is below $10^7\,\Msun$
this is well above the maximum accretion rate permitted by the Eddington limit
which is -- assuming an accretion efficiency of $10\%$ -- of the order of 
$\dot{M}_\mathrm{max}\approx 2\cdot10^{-8}(M/\Msun)\,\Msun\,\yr^{-1}$.
Nevertheless, if one assumes that the black hole accretes at the Eddington limit
during its early evolution switching to the constant sub-Eddington accretion rate
after the Salpeter time scale \citep{salpeter1964}, it is still possible to
accumulate up to $10^9\,\mathrm{\Msun}$ within a few billion years \citep{duschl2011}.

\begin{table}
\centering
\begin{tabular}{ccccc}
\toprule
\makebox[4em]{$\tilde{r}$} & \makebox[4em]{$\tilde{\tau}$} & \makebox[3em]{$\tilde{t}$} &
  \makebox[3em]{$\widetilde{\Omega}$} & \makebox[3em]{$\tilde{v}$} \\
\midrule
 $1\,$AU & $5\cdot10^{-3}\,$yr & $5\,$yr & $200\,$yr$^{-1}$ & $948\,$km\,s$^{-1}$ \\
\bottomrule
\end{tabular}
\medskip
\centering
\begin{tabular}{cccc}
\makebox[3em]{$\widetilde{M}$} & \makebox[3em]{$\widetilde{\sden}$} & \makebox[3em]{$\widetilde{\dot{M}}$} &
\makebox[3em]{$\widetilde{Q}$}\\
\midrule
$10^3\,$M$_\odot$ & $9\cdot10^9$ g\,cm$^{-2}$ & $200\,$M$_\odot\,$yr$^{-1}$ & $5\cdot10^{17}\,$W\,m$^{-2}$ \\
\bottomrule
\end{tabular}
\caption{Scales for the AGN example with $\beta=10^{-3}$. Most of the values have
been rounded to one significant digit to fit into the table. \alert{For typical parameters
of an evolved AGN disc see \citet{collin1990,lin1996}.}}
\label{tab:scaling_AGN}
\end{table}

If one assumes that energy transfer inside the disc is mainly due to radiation
in the vertical direction one can estimate the effective temperature of the disc
by equating the energy generation due to viscous dissipation and energy loss
caused by radiative cooling:
\begin{equation}
  Q_\mathrm{vis}=Q_\mathrm{cool}=2\sigma\,T_\mathrm{eff}^4
\end{equation}
where $\sigma$ is the Stefan-Boltzmann constant. The factor $2$ is necessary to
account for both radiating surfaces of the disc. Since $Q_\mathrm{vis}$ is
completely determined by the self-similar solution (see
Eq.~\ref{eqn:dissipation_rate_nondim}), we can solve the equation
above for $T_\mathrm{eff}$.

\begin{figure}
  \centering
  \includegraphics[width=\linewidth]{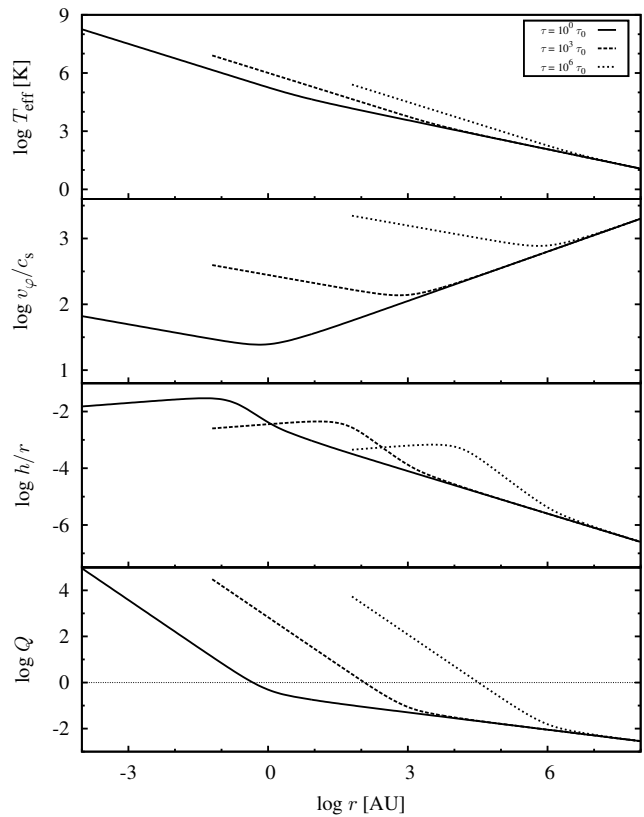}
  \caption{Self-similar time evolution of a massive AGN disc. The diagrams show
    effective temperature $T_\textnormal{eff}$, Mach number $\vphi/\cs$, scale
    height $h$ and Toomre parameter $Q$ as a function of radius at three different
    times. The results were obtained with the same parameters as those depicted
    in Figs.~\ref{fig:time_evolution_dsb} and~\ref{fig:time_evolution_dsb1},
    namely zero torque boundary condition with DSB viscosity prescription and
    parameters $\kappa=-1, \beta=10^{-3}, \tau_0=1, \Mc(\tau_0)=1,\ln\xi_0=-10$.
    The solutions are truncated at $3$ Schwarzschild radii (see text).}
  \label{fig:time_evolution_dsb_AGN}
\end{figure}

The result for the AGN example is shown in Fig.~\ref{fig:time_evolution_dsb_AGN}
(upper panel). In principle we can compute the temperature distribution for
arbitrary small radii, but below a certain minimal radius it becomes meaningless.
Therefore we decided to truncate the curves at the last stable circular orbit
around a Schwarzschild black hole located at three Schwarzschild radii
\citep{novikov1973}. In view of the previously shown solutions, it is not
astonishing that the radial temperature profile is given by a broken power law.
The exponents are determined by the exponents of $Q_\mathrm{vis}$ (see
Tab.~\ref{tab:asymptotic_exponents_radial}) multiplied by a factor of $1/4$ which
becomes $-3/4$ at small radii in case of the zero torque boundary condition and
$(5\kappa+3)/4=-1/2$ (with $\kappa=-1$) in the self-gravitating outer regions of
the disc. The slope of the radial temperature profile, say $p$, is related to the
slope of the spectral energy distribution according to $n=3-2/p$
\citep[see][]{lyndenbell1969}. Again our model recovers the spectral index $n=1/3$
for Keplerian discs, but in the self-gravitating region this becomes $n=-1$
leading to an enhanced radiative flux at higher wave lengths.

So far we have avoided the intricate computation of the vertical disc structure.
This would in general require the solution of the energy equation in conjunction
with the radiative transfer equation which is beyond the scope of this work.
However, if we assume an isothermal vertical structure -- which is a very gross
simplification -- the midplane temperature equals the effective temperature. This
allows us to estimate the speed of sound and the azimuthal Mach number $\vphi/\cs$
in the equatorial plane depicted in the second panel of Fig.~\ref{fig:time_evolution_dsb_AGN}.
The results show that the rotational velocity is indeed supersonic which is 
required by the model assumptions (see Sec.~\ref{sec:supersonic_rotation}).
In the early phases of the AGN evolution the flow is only slightly supersonic
with a minimum Mach number of the order of $10$ around the region between the
Keplerian and the self-gravitating disc. The minimum remains in this region
during the evolution of the AGN, but its value increases up to roughly $10^3$.

With the estimate of the azimuthal Mach number one can verify another important
requirement for the model, namely the thin disc assumption $h/r\ll 1$. This
ratio of scale height and radius is shown in the third diagram of
Fig.~\ref{fig:time_evolution_dsb_AGN}. Since this ratio is related to the
azimuthal Mach number according to Eq.~(\ref{eqn:supersonic_rotation_fsg}),
the assumption is again better for the evolved AGN disc. However, it remains
always smaller than one but reaches values of the order of $10^{-2}$ in the
early phases of the evolution.

\alert{Although our cooling model is very simple, our results on effective 
temperature, Mach number and aspect ratio are in good agreement with those given
in the literature \citep[see \eg][]{collin1990,lin1996}, at least in case of
an evolved AGN disc. Besides that one should be aware of the fact that any
results on the early evolution of an AGN disc are quite speculative, because
observational constraints on accretion discs around intermediate mass black holes
with masses in the range of $10^3$ to $10^5\,\Msun$ are very limited. Right now
there seems to be one promising candidate \citep{davis2011}.}

Another important quantity when talking about self-gravitating accretion discs
is the Toomre stability parameter $Q$ defined in Eq.~(\ref{eqn:toomre_paramter}).
If its value is below one, the disc becomes gravitationally unstable. Keplerian
discs are always stable. This can be seen most easily with help of Eq.~(\ref{eqn:self_regulation}).
If the logarithmic gradient $x$ of the rotational velocity approaches the
Keplerian value $Q$ becomes infinitely large as long as the inverse of the
azimuthal Mach number $\cs/\vphi$ remains finite. On the other hand, if $x$
deviates from its Keplerian value of $-3/2$ the disc may become unstable
if the rotational velocity is highly supersonic. In Fig.~\ref{fig:time_evolution_dsb_AGN}
(lower panel) the Toomre parameter for the AGN disc is shown as a function of
radius at three different times. The outer regions of the disc are always unstable,
because the slope of the rotational velocity deviates considerably from its
Keplerian value and the Mach number becomes rather high. In the early phases of
the evolution the transition from stable to unstable occurs roughly in the 
intermediate region where the disc becomes fully self-gravitating. This location
is shifted towards lower radii as time progresses, because the Mach numbers in
the evolved discs are higher even in the region where the discs are almost
Keplerian.

One might argue that the existence of these solutions is questionable, because
the gravitational instabilities would lead to fragmentation of the whole disc.
Since the time scales for these local gravitational collapses are rather short
compared to the viscous time scale, the disc would disintegrate into small clumps
forming stars before a considerable amount of gas has been accreted on to the
central black hole \citep{lin1987,shlosman1990}. Since these processes are
very sensitive to the temperature of the disc \alert{\citep{gammie2001}} one
should be careful with this reasoning, because we applied a far too simple cooling model to derive the
temperature distribution. As was already mentioned above, the deduction of a
reliable disc temperature is quite difficult and beyond the scope of this paper.
Therefore we leave it for future work.

\begin{alertenv}
\subsection{Comparison with other works}
\label{sec:comparison}

The literature on one-dimensional modelling of self-gravitating accretions
discs using the local viscous transport approximation can be subdivided into
two different categories: (i) Keplerian models in which the rotation law is not
affected by self-gravity and (ii) fully self-gravitating disc models in which
one considers the discs potential in the radial balance law. In both cases there
exist stationary as well as non-stationary models. We briefly discuss the
former here, because they are not subject of the present work, and compare our
results in more detail with other time-dependent models.

\cite{paczynski1978} proposed the first self-gravitating AGN model considering
an evolved AGN with supermassive black hole mass of $10^{10}\,\Msun$ surrounded
by a rather compact Keplerian disc ($M_\mathrm{disc}=10^9\,\Msun$). The surface
density in this model shows a steep decline ($\propto r^{-3}$), which we never
yield in our models due to the constraints on $\kappa$ (see
Tab.~\ref{tab:asymptotic_exponents_radial}). Other authors
\citep{mineshige1996,bertin1997,bertin1999,duschl2000} find fully
self-gravitating stationary solutions with different viscosity prescriptions
including $\alpha$-viscosity (with self-regulation) and $\beta$-viscosity. All
these models show a remarkable common feature, namely that the rotation curves
become flat and that $\sden\propto r^{-1}$ in the limit $r\to\infty$. We find
the same asymptotic behaviour and an almost constant accretion rate in our
self-similar quasi-stationary model which we obtain for $\kappa=-1$ (see
Sec.~\ref{sec:time_evolution}).

\cite{lin1987} developed a time-dependent self-similar model for self-regulated
Keplerian discs accounting for self-gravity only in their viscosity model (see
Sec.~\ref{sec:viscosity_prescription}). They found an analytic solution for
the zero torque boundary condition if total disc angular momentum is conserved.
Within the context of our self-similar model we already showed that the case of
constant disc angular momentum is not permitted by the basic constraints on the
slope of the rotation curve, because it would require that $\kappa=-5/3$ (see
Sec.~\ref{sec:conservation_conditions}). The analysis of the phase plane for our
model reveals that in this particular case the local exponent of the rotation
law $x$ becomes smaller than the Keplerian value of $-3/2$ at some finite radius
for all solutions which are Keplerian near the origin. Thus beyond this radius
the slope of the rotation law is steeper than permitted by physical reasons and
as a consequence $\sden$ becomes negative (see Eq.~\ref{eqn:sigma_omega_relation}).
Interestingly the solution of \cite{lin1987} yields complex values for $\sden$
beyond a certain radius. Hence this solution also breaks down at the outer
rim of the disc. Apart from these general problems our results exhibit some
features of their solution near the origin, namely that $\sden\propto r^{-3/2}$
for all solutions obtained for LP-viscosity with no central coupling (see
Tab.~\ref{tab:asymptotic_exponents_radial}) and that $\dot{M}\propto t^{-6/5}$
for $\kappa=-5/3$ (see Tab.~\ref{tab:asymptotic_exponents_temporal}).

\cite{rice2009} extended this model considering a cooling mechanism which
allows to determine the $\alpha$ parameter of the effective viscosity
self-consistently \cite[see][]{gammie2001}. Although the authors claim to solve
a fully self-gravitating disc model, we would prefer the term \emph{Keplerian model}
here, because they do neither account for the disc material in the radial
balance law (Eq.~\ref{eqn:centrifugal_gravity_balance}) nor do they consider
the impact of self-gravity on the scale height (see Sec.~\ref{sec:vertical_structure}
and Eq.~\ref{eqn:supersonic_rotation_fsg}). Nevertheless, we can compare their
numerical solutions for the time evolution of the surface density with our
self-similar model. A remarkable common feature is again that $\sden\propto r^{-3/2}$
for small radii corresponding to our model with LP-viscosity and zero torque
boundary condition. At about $2\,\mathrm{AU}$ the slope of the surface density
becomes steeper ($\sden\propto r^{-2}$) which is the asymptotic limit as $r\to\infty$
for almost massless discs ($\kappa \approx -3/2$) in our self-similar model 
(see Tab.~\ref{tab:asymptotic_exponents_radial}). Then at about $20\,\mathrm{AU}$
the decline in surface density steepens again. According to the authors this is
due to their cooling model which causes a sudden drop in temperature. Since our
model does not account for any cooling mechanism one cannot expect to reproduce
this feature.

Based on their previous work \cite{lin1990} extended their model accounting
for self-gravity (in the monopole approximation), radiative cooling and energy
transport. Furthermore they modified the viscosity prescription adding
the standard Shakura-Sunyaev viscosity to the original model \citep{lin1987}
to avoid the problem of a vanishing viscosity coefficient in the inner Keplerian
parts of the disc. They performed numerical simulations for different initial
conditions. The radial profiles of surface density evolve self-similar in all
their simulations with power law exponents between $-6/5$ and $-4/3$
(depending on the model) in the fully self-gravitating region. It flattens close
to the inner boundary of the disc where the Shakura-Sunyaev viscosity dominates.
This is consistent with our observation that $\sden\propto r^{-1/2}$ for $r\to0$
with DSB $\beta$-viscosity and zero torque boundary condition (see
Tab.~\ref{tab:asymptotic_exponents_radial}). We already showed in
Sec.~\ref{sec:viscosity_prescription} that $\beta$-viscosity recovers
$\alpha$-viscosity in this limit.

Furthermore we find a remarkable match comparing the time evolution of the
central mass with our self-similar model. If one reads off from their diagrams
the power law exponents of the radial surface density distribution (at a
specific time!), say $\sigma$, one can compute the associated exponents of the
rotation law according to $\kappa=(\sigma-1)/2$ (see last column of
Tab.~\ref{tab:asymptotic_exponents_radial}). With these values for $\kappa$ one
can make a prediction for the time evolution of the central mass using the
asymptotic exponent given in the third row of Tab.~\ref{tab:asymptotic_exponents_temporal}.
From their model A1, for instance, we get $\sigma=-6/5$ and therefore $\kappa=-11/10$.
Thus we predict that $\Mc\propto t^{8/11}$. This is almost exactly the value one
can read off their diagram ($\approx0.73$) for the period in which $\sden$ evolves
quasi-stationary with a constant radial slope (between $t=2\,t_\mathrm{ff}$ and
$3\,t_\mathrm{ff}$). The exponent for four different numerical models and our
predictions are shown in Tab.~\ref{tab:comparison_exponents}. This strongly
supports our observation that the self-similar time evolution of self-gravitating
discs is mainly controlled by a single dimensionless parameter, namely $\kappa$.

\begin{table}
\centering
\begin{alertenv}
\begin{tabular}{ccccc}
\toprule
LP-model & $\sigma$ & $\kappa$ & $\mu$ & $-\left(3/\kappa+2\right)$\\
\midrule\smallskip
A1 & $-\tfrac{6}{5}$ & $-\tfrac{11}{10}$ & $0.73$ & $\tfrac{8}{11}\approx0.73$ \\\smallskip
B1 & $-\tfrac{5}{4}$ & $-\tfrac{9}{8}$ & $0.67$ & $\tfrac{2}{3}\approx0.68$ \\\smallskip
A4 & $-\tfrac{13}{10}$ & $-\tfrac{23}{20}$ & $0.61$ & $\tfrac{14}{23}\approx0.61$ \\\smallskip
B4 & $-\tfrac{4}{3}$ & $-\tfrac{7}{6}$ & $0.58$ & $\tfrac{4}{7}\approx0.57$ \\
\bottomrule
\end{tabular}
\caption{Comparison of radial and temporal power law exponents of surface
 density $\sden\propto r^\sigma$ and central mass $\Mc\propto t^\mu$ derived from
 the self-similar model with values obtained from \protect\cite{lin1990}
 (columns 2 and 4).}
\label{tab:comparison_exponents}
\end{alertenv}
\end{table}

Fully self-gravitating self-similar accretion disc models were first developed
in \cite{mineshige1997} and \cite{mineshige1997a}. In the first paper the authors
assume that the disc is isothermal with respect to $r$ whereas in the second
paper they extend the model using a polytropic relation. Both papers account for
self-gravity in the monopole approximation and simplify the equations using the
slow accretion limit. The main difference compared to our model is the effective
viscosity which depends linearly on radius or -- in the second paper --
according to a power law with a constant exponent. Thus the viscosity does not
change as the disc evolves in time. Another important difference of these models
is the similarity transformation applied to the system of differential
equations. In \cite{mineshige1997} the similarity variable is proportional to
$r/t$ which corresponds to our quasi-stationary model obtained with $\kappa=-1$.
At large radii the asymptotic behaviour of these similarity solutions is in
compliance with our findings ($\sden\propto r^{-1}$, $\Omega\propto r^{-1}$, 
$\dot{M}=\mathsf{const}$, $\vs=\mathsf{const}$). Whereas for small radii the
model fails, because the rotation law does not become Keplerian, instead 
$\Omega\propto r^{-4/3}$. Furthermore the accretion rate is proportional
to $r^{1/3}$ in the vicinity of the central object. It therefore vanishes in
the limit $r\to 0$. The results for the polytropic model do not differ
substantially from the isothermal model and suffer from the same limitations.

On the basis of the isothermal model presented in \cite{mineshige1997}
\cite{tsuribe1999} developed an isothermal self-similar $\alpha$-disc solution
depending on a variable Toomre parameter. He utilized the viscosity prescription
in Eq.~(\ref{eqn:alpha_viscosity_general}) together with the definition of the
Toomre parameter (Eq.~\ref{eqn:toomre_paramter}). This allows him to express the effective
viscosity in terms of $\Omega$ and $\sden$ in the same way as we do to derive
the LP-viscosity model (Eq.~\ref{eqn:beta_viscosity_LP}). The only difference
between both viscosity prescriptions is that we additionally assume that
the disc is in a marginally stable state, \ie $Q=1$. Thus his similarity
solutions depend on the parameter $Q$. Apart from this, his model is identical
to our quasi-stationary model for LP-viscosity and zero torque boundary condition
with $\kappa=-1$. The comparison with our results reveals that all observables
show exactly the same asymptotic behaviour (see Tab.~\ref{tab:asymptotic_exponents_radial}
and \ref{tab:asymptotic_exponents_temporal}). Hence, this solution is a special
case of our more general results.

More recently \cite{abbassi2006,abbassi2013} proposed self-similar disc models
using the $\beta$-viscosity prescription. As was already pointed out in the
introduction of the present work, these models are based on contradictory
assumptions. Thus consequently the solutions presented in these works violate
a fundamental requirement for the applicability of the slow accretion limit:
highly supersonic azimuthal velocity $\vphi \gg \cs$ (see Fig.~2 of
\cite{abbassi2006}). Hence, we do not comment further on their results.
\end{alertenv}

\section{Summary}
\label{sec:summary}

In this work we have developed a rather comprehensive dynamical model for
geometrically thin accretion discs. It accounts for time-dependent angular
momentum redistribution and mass accretion on to the central object as well as
the impact of self-gravity on the accretion process. Several reasonable
assumptions have been made to reduce the complexity of the problem keeping the
model as simple as possible including the thin disc approximation, the slow
accretion limit and the monopole approximation. On the basis of these
assumptions we have deduced a new PDE describing the time evolution  of the
angular velocity profile in viscous, self-gravitating accretion discs. This can
be seen as the major difference between our approach and the standard theory
where disc dynamics is based on the evolution of surface density. However,
since angular velocity is related to the mass distribution we can derive the
surface density from our solutions which makes our model coequal, albeit more
simple in view of the underlying mathematical problem.

We have discussed three different well-established viscosity models taken from
the literature, all of them have already been applied to self-gravitating
accretion discs. It is worthwhile to mention here that although one has to
specify a viscosity prescription in order to solve the disc evolution equation,
our model is far more general in the first place, because it does not rely on a
specific viscosity model. We have furthermore shown that in case of the three
viscosity prescriptions the disc evolution equation admits similarity solutions.
The associated similarity transformation depends on a single non-dimensional
parameter which is related to the radial slope of the angular velocity at large
radii. We have shown that this parameter has a major impact on the whole disc
evolution and that flatter rotation laws yield higher accretion rates. Thus
we conclude that the more self-gravitating the accretion disc, the faster the
growth of the central object.

Finally, we have applied our model to an AGN accretion disc. We have discussed
the formation of supermassive black holes in this context and found that our 
model of self-gravitating accretion discs may explain the rapid growth of SMBHs
in the early universe.

%% list of references using bibtex
\bibliography{ssdisk}

%% appendices
\begin{appendix}

\section{The functions $\mathcal{F}_<$ and $\mathcal{F}_>$}
\label{app:weight_functions}

\begin{figure}
  \centering
  \includegraphics[width=\linewidth]{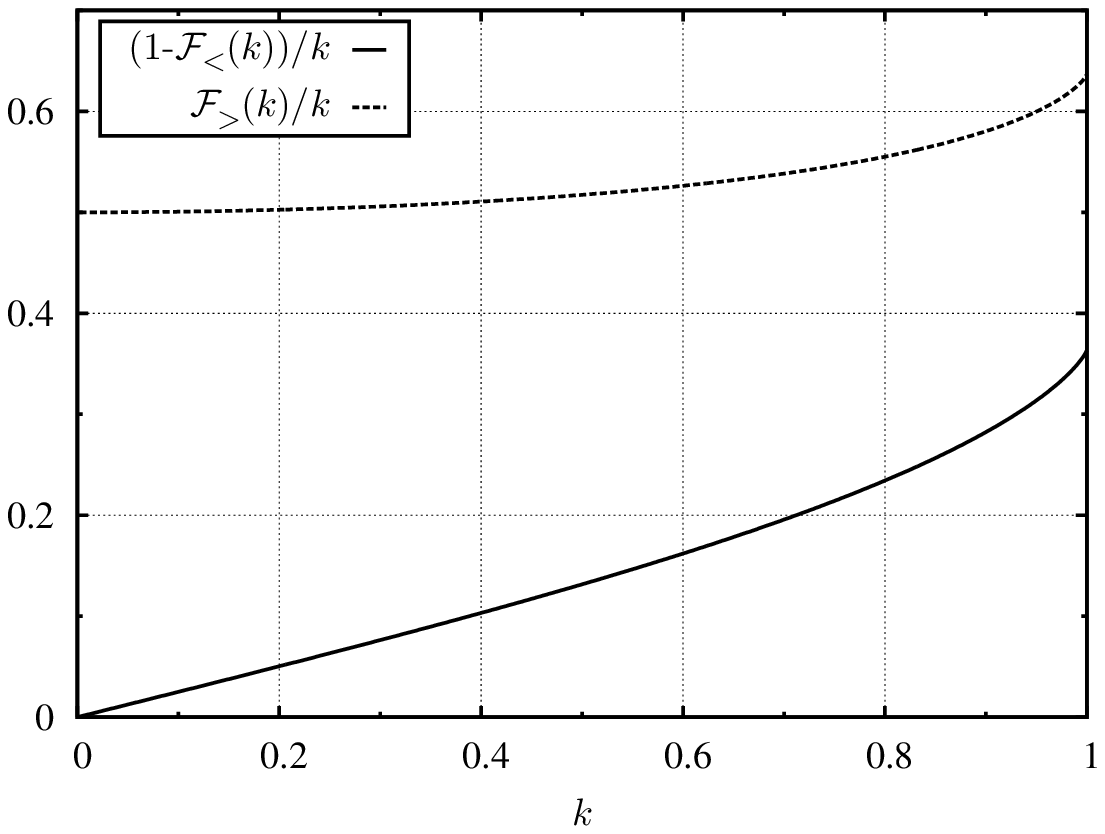}
  \caption{Dimensionless weight functions}
  \label{fig:weight_functions}
\end{figure}

The definite integral in Eq.~(\ref{eqn:Fltgt_functions}) exists for any
combination of the parameters $r\geq0$ and $s\geq0$ where at least one of
them is different from zero. An analytical expression in terms of hypergeometric
functions is given in \cite{gradshteyn2000} (Eq.~6.574). It may be rewritten
with help of the complete elliptic integrals of the first and second kind $K(k)$
and $E(k)$ according to
\begin{align*}
		\mathcal{F}_<(k)&=\frac{2}{\pi}E(k)&
		&\mathrm{if}& k&=\tfrac{s}{r}\leq1 \\
		\mathcal{F}_>(k)&=\frac{2}{\pi}\frac{E(k)-(1-k^2)K(k)}{k}&
		&\mathrm{if}& k&=\tfrac{r}{s}<1
\end{align*}
where $k$ is the elliptic modulus. One easily proves using the asymptotic
expansions of $E(k)$ and $K(k)$ that
\begin{equation*}
  \mathcal{F}_<(k) = 1-\bigl(\tfrac{k}{2}\bigr)^2+\order{k^4}
  \quad\mathrm{and}\quad
  \mathcal{F}_>(k) = \tfrac{k}{2} + \order{k^3}.
\end{equation*}
In order to compute the integral expressions in Eq.~(\ref{eqn:poisson_solution2})
one has to evaluate the weight functions $(1-\mathcal{F}_<)/k$ and
$\mathcal{F}_>/k$ plotted in Fig.~\ref{fig:weight_functions}.

\end{appendix}

\clearpage

\end{document}